\documentclass[aps,prl,reprint,balancelastpage,showpacs,floatfix]{revtex4-1}
\usepackage{amsmath,amssymb}
\usepackage{graphicx}
\usepackage{braket}

\begin{document}

\onecolumngrid
\title{Observation of Chiral-Mode Domains in a Frustrated XY Model on Optical Triangular Lattices}
\author{Hideki Ozawa}
\altaffiliation{Electronic address: hideki.ozawa@riken.jp}
\affiliation{RIKEN Center for Quantum Computing (RQC), Wako 351-0198, Japan}
\author{Ryuta Yamamoto}
\affiliation{RIKEN Center for Quantum Computing (RQC), Wako 351-0198, Japan}
\author{Takeshi Fukuhara}
\affiliation{RIKEN Center for Quantum Computing (RQC), Wako 351-0198, Japan}
\date{\today}

\begin{abstract}
We investigated the relaxation and excitation in a frustrated XY model realized by a Bose gas in Floquet-engineered optical triangular lattices. Periodically driving the position of the entire lattice structure enables the sign inversion of tunneling amplitudes, which, in the case of a triangular lattice, results in geometrical frustration of the local phase of wave packets. We revealed that the two spiral phases with chiral modes show significant differences in relaxation time from the initial ferromagnetic phase. While spontaneous symmetry breaking is clearly observed at a slow ramp of the Floquet drive, simultaneous occupation of two ground states often occurs at a fast ramp, which can be attributed to the domain formation of the chiral modes. The interference of the spatially separated chiral modes was observed, using a quantum gas microscope. This work leads to exploring the domain formation mechanism in a system with U(1)$\times \mathbb{Z}_2$ symmetry.
\end{abstract}
\maketitle
%
%
Magnetic frustration is an intriguing issue in condensed matter physics~\cite{Diep:2004,Moessner:2006}. The simplest example is spins with antiferromagnetic interactions in a triangular lattice, in which all adjacent spins cannot align in antiparallel configurations that minimize the interaction energy. Owing to geometrical frustration, conventional magnetic orders are suppressed, giving rise to non-trivial phenomena and phases such as quantum spin liquids~\cite{Balents:2010}. However, theoretical challenges remain especially for quantum spin systems. In the experimental side, conventional condensed matter systems are too complex to realize ideal models of frustrated spin systems~\cite{HarrisonA2004}. Quantum simulators, controllable physical systems that realize target models, including frustrated spin models, are expected to play a significant role in understanding frustration physics. Such studies have been conducted using various platforms including trapped ions~\cite{Kim:2010,Qiao:2022}, neutral atoms in optical lattices~\cite{Struck:2011,Mongkolkiattichai:2022,XuM2023,LebratM2023,PrichardM2023} and in optical tweezer arrays~\cite{Scholl:2021,Semeghini:2021}, superconducting annealers~\cite{King:2021}, and Josephson junction arrays~\cite{Cosmic:2020}. 

Two-dimensional fully frustrated XY models, such as the antiferromagnetic XY model on a triangular lattice, have attracted attention in the last decades~\cite{Teitel:1983,Miyashita:1984, Lee:1984, Song:2022}. The main feature of the models is the discrete $\mathbb{Z}_2$ symmetry stemming from two-fold degenerate ground states corresponding to the two chiral modes. There have been many controversial discussions on classical spin models because the phase transition associated with the chiral $\mathbb{Z}_2$ symmetry breaking occurs at a temperature very close to the transition temperature corresponding to the breaking of the continuous U(1) symmetry for global spin rotation. Although the transition temperature of the chiral symmetry breaking is slightly higher than the other transition temperature, there is still no clear consensus on the critical behaviors of transitions~\cite{Obuchi:2012,Lv:2013}. For quantum spin models, this combined U(1)$\times \mathbb{Z}_2$ degeneracy might bring about exotic quantum critical phenomena; however, such critical behaviors have not been elucidated. 

Quantum simulation of classical XY spin model has been demonstrated by using ultracold bosonic atoms in an optical triangular lattice~\cite{Struck:2011}. To realize each ground state in the model, the tunneling amplitudes were manipulated by the lattice shaking technique~\cite{RevModPhys.89.011004}. While the interference patterns of the ground states have been observed, relaxation and excitation from the initial ferromagnetic state have rarely been studied.

\begin{figure*}[!tb]
\begin{center}
\includegraphics[width=160mm]{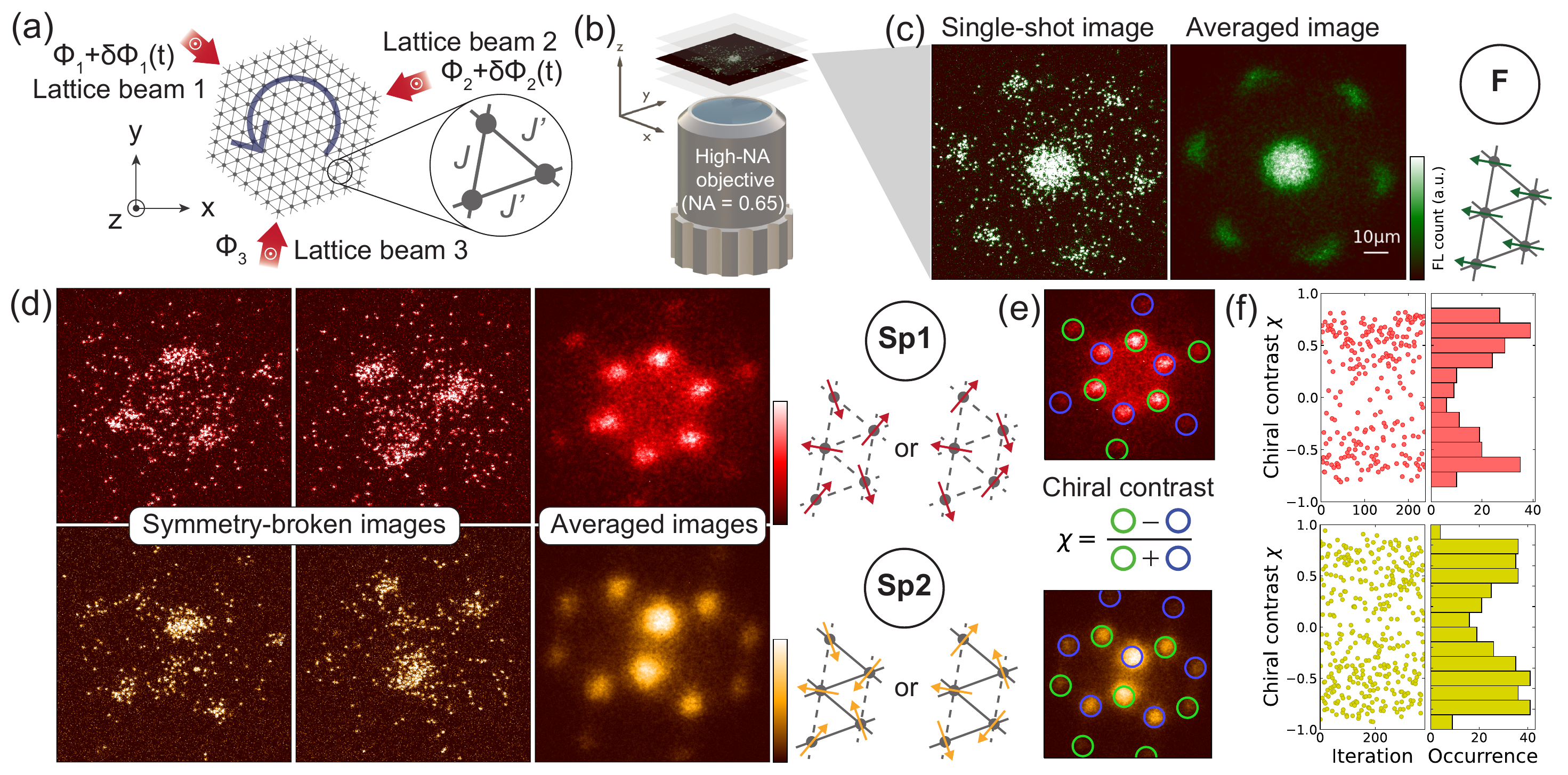}
\caption{Schematic of shaken optical triangular lattice and realization of XY spin model. (a) The tunneling parameters $J$ and $J'$ can be tuned independently by modulating the phase of two lattice beams. (b) Observation with the quantum gas microscope. The samples are loaded into a vertical lattice, and the single layer located at the microscope's focus is selectively detected by removing atoms in the other layers before the measurement. (c) Observation of a ferromagnetic phase~(F). Color bars indicate fluorescent (FL) counts. Arrows on the rightmost sketch mean the spin state. (d) Observation of spontaneous symmetry breaking in two phases with frustration: Spiral~1~(Sp1) and Spiral~2~(Sp2). The two columns on the left show the single-shot symmetry-broken images with different chiral modes. For the averaged images in (c) and (d), 20 and 100 independent experimental realizations were used, respectively. The solid (dashed) lines in the rightmost column mean tunnelings with positive (negative) signs. (e) Definition of chiral contrast $\chi$, which indicates chiral order. (f) Statistical distributions (left) and histograms (right) of $\chi$ for Sp1 and Sp2.}
\label{image1}
\end{center}
\end{figure*}

In this study, we focused on this aspect. The tunneling amplitudes $J,J'$ in the optical triangular lattice are independently controlled by modulating two phases $\phi_1,\phi_2$ of the three lattice beams (Fig.~\ref{image1}(a)). By varying the time to ramp up the phase modulation amplitudes, we investigated and compared the relaxation times from the initial ferromagnetic phase (F) to two frustrated phases (Sp1, Sp2). We combined the lattice shaking technique with a quantum gas microscope, which has a single-site resolution and single-atom sensitivity (Fig.~\ref{image1}(b)). This experimental system is capable of investigating phase separation and density waves arising from exotic phases such as lattice supersolidity~\cite{Wang:2009,Jian:2009,Heidarian:2010}.

%
%

First, we describe our experimental setup. A sample was prepared by loading a Bose-Einstein condenstate~(BEC) of $^{87}$Rb atoms into a lattice system consisting of an optical triangular lattice and crossed far-off resonance traps (FORT). The optical triangular lattice potential is given by:
\begin{eqnarray}
V (\boldsymbol{r}) &=& - \frac{V_0}{2} \left[ {\rm cos}(\boldsymbol{b}_1 \cdot \boldsymbol{r}+\phi_{23}) + {\rm cos}(\boldsymbol{b}_2 \cdot \boldsymbol{r}+\phi_{31}) \right. \nonumber \\
&& \left.  + {\rm cos}(\boldsymbol{b}_3 \cdot \boldsymbol{r}+\phi_{12})  \right]+\frac{1}{2}m \omega_z^2 z^2, \label{eq:TriLPotential}
\end{eqnarray}
where $V_0$ is the lattice depth, $\boldsymbol{b}_i$ the reciprocal lattice vectors, $\omega_z/2\pi$ the harmonic trap frequency along the direction perpendicular to the lattice plane, and $\phi_{ij} = \phi_i - \phi_j$ is the relative phase between two of the three lattice beams, for which we choose the wavelength $\lambda=1064$ nm. In Eq.~\ref{eq:TriLPotential}, we omit the offset term and the influence of the external trap frequencies in the xy-plane for simplicity. Unless otherwise mentioned, the atoms were initially loaded to a lattice depth of $V_0=3.0~E_R$, where $E_{R}= \hbar^2 k_{L}^2/2m$ is the recoil energy, $k_{L}=2 \pi/\lambda$ the wave number, $\hbar$ the Planck constant divided by $2\pi$, and $m$ the mass of $^{87}$Rb atom. The Hubbard parameters are $U/h=30.7$ Hz, $J_{\rm bare}/h=26.9$ Hz, where $U, J_{\rm bare}$ are the on-site interaction and the nearest neighbor tunneling, respectively. The external trap frequencies are $(\omega_x, \omega_y, \omega_z)/2\pi =(88,150,184)$ Hz.

%
%

After lattice loading, we increased phase modulation signals to shake the optical triangular lattice elliptically (Fig.~\ref{image1}(a)). According to the Floquet theory, the effective tunnelings $J$ and $J'$ in the rotating frame obey the zeroth order Bessel function of the first kind (see Supplementary Material for the details of lattice shaking parameters and experimental sequence~\cite{supp_Lieb}). The effective tunnelings are $(J,J')/J_{\rm bare}=(-0.35,-0.35)$ for Sp1, and $(-0.35,0.35)$ for Sp2 throughout this letter. The modulation frequency $\Omega/2\pi=1.2$ kHz was carefully chosen to avoid multi-photon interband excitations~\cite{Weinberg:2015,supp_Lieb}. We also note that the crossed FORTs depths after lattice loading had to be lowered as much as possible so that the evaporation of atoms heated by lattice shaking could work well~\cite{PhysRevLett.119.200402,supp_Lieb}. We observed the interference patterns of atoms using in-plane time-of-flight (TOF)~\cite{Bakr:2010}, where the triangular lattice potential was suddenly switched off, whereas the vertical lattice potential was ramped up so that the atomic cloud could expand within the layers. Atoms were typically split into 3 layers in the vertical lattice, and more than 60$\%$ of the atoms were populated in a target layer of imaging. The in-plane TOF was followed by (i) a sudden ramp-up of all the optical lattices to freeze out atoms, (ii) selection of the target layer by the combination of microwave and B-field gradient, and (iii) fluorescence imaging using the Raman sideband cooling~\cite{RYamamoto:2020}. Figures~\ref{image1}(c) and (d) show single-shot and averaged images of atom distributions after 5 ms in-plane TOF. In the case of Sp1 and Sp2, where two-fold chiral degeneracy exists, the symmetry-broken images are observed~\cite{Struck:2011}. We also checked the statistical distribution of the chiral contrast $\chi$, which is defined in Fig.~\ref{image1}(e). The histograms appear binary, indicating that symmetry breaking often happens. We took the data at a ramp-up time of 200 ms for Sp1 and 300 ms for Sp2 with crossed FORT depth optimized for each frustrated phase~\cite{supp_Lieb}. The atom number $n$ per tube in the shaken lattice differed in each phase because the loss rate during lattice shaking depends on the phase modulation amplitudes. For example, in Sp2, the filling was $n\sim6$. Since $U/n|J^{( \prime )}|\sim 0.5$, this experiment is conducted in the weak interaction regime, where the system is mapped to the classical XY model~\cite{Eckardt:2010}.

%
%

\begin{figure}[!tb]
\begin{center}
\includegraphics[width=80mm]{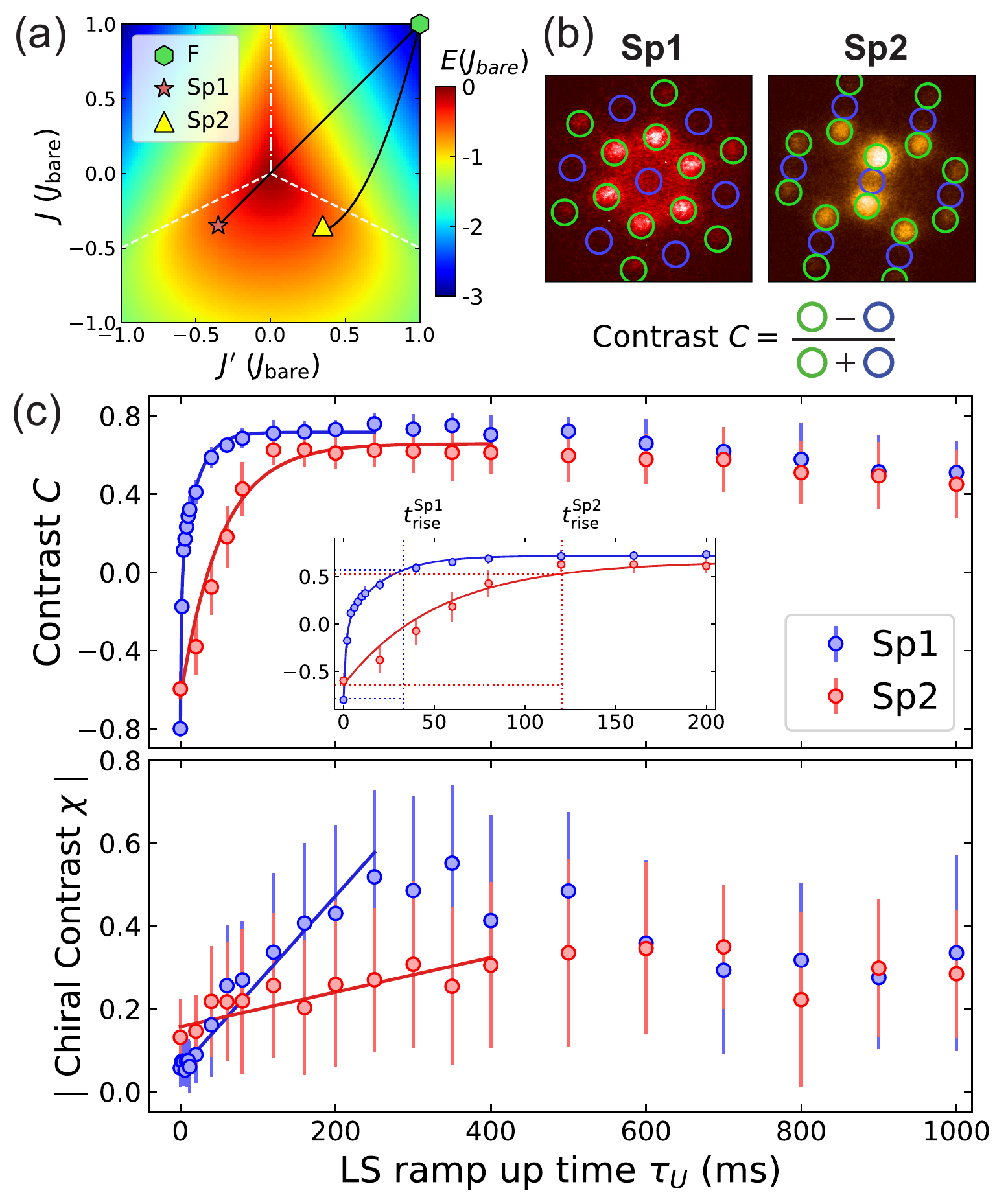}
\caption{Relaxation from F to Sp1 or Sp2. (a) Phase diagram of the classical XY model. The white (dot-)dashed line means the phase transition of the 1st(2nd) order. The black solid lines represent the paths from F to Sp1 or Sp2. (b) Definition of contrast $C$. (c) $C$ and $|\chi|$ with various ramp-up times of lattice shaking $\tau_{\rm U}$. Blue (red) points represent Sp1 (Sp2) data. Error bars denote standard deviations. The solid lines are fitting results to the data. The fitting function for $C$ is defined in Eq.~\ref{eq:FitFuncC}, while that for $|\chi|$ is a linear function with offset, respectively. The inset shows $C$ from 0 ms to 200 ms ramp-up time together with the extracted rise times~$t_{\rm rise}^{\rm Sp1(2)}$.}
\label{image2}
\end{center}
\end{figure}

In the following, we focus on the relaxation from F to Sp1 and Sp2 (see Fig.~\ref{image2}(a)). 
To quantify the relaxation times, we introduce contrast $C$ defined in Fig.~\ref{image2}(b). When $C<0$, the system is in an F state. $C>0$ indicates a phase transition to Sp1 or Sp2. Figure~\ref{image2}(c) shows $C$ and $|\chi|$ with various ramp-up times of the phase modulation signals. The relaxation time of Sp1 is much shorter than that of Sp2. We made a fit to the data of $C$ with our empirical exponential functions
\begin{eqnarray}
f(t) =  \left\{ 
    \begin{array}{ll}
    \frac{A}{2} \left( e^{-t/\tau_{\rm fast}} + e^{-t/\tau_{\rm slow}} \right) + B & {\rm for} \; {\rm Sp1} \\
    A e^{-t/\tau} + B & {\rm for} \; {\rm Sp2}, \\
    \end{array}
\right.
\label{eq:FitFuncC}
\end{eqnarray}
where $A, B, \tau, \tau_{\rm slow}, \tau_{\rm fast}$ are fitting parameters. We define the rise time $t_{\rm rise}^{\rm Sp1(2)}$ such that $f\left(t_{\rm rise}^{\rm Sp1(2)}\right) = 0.1 A + B$
is satisfied. The extracted rise times are $t_{\rm rise}^{\rm Sp1}=32.9$ ms, $t_{\rm rise}^{\rm Sp2}=120$ ms. We attribute this difference to two factors; one is the path length ratio after the phase transition over the total length (see black solid lines in Fig.~\ref{image2}(a)). The ratio for Sp1 is 2.6~times larger than that for Sp12. The other is the effective band structures. In the tight-binding approximation, the energy difference between the ground states of Sp1 and $\Gamma$ point, at which a BEC is populated before lattice shaking, amounts to $\Delta E_{\Gamma}^{\rm Sp1}=3.6J_{\rm bare}$; on the other hand, the counterpart of Sp2 is $\Delta E_{\Gamma}^{\rm Sp2}=0.38J_{\rm bare}$, 9.5 times smaller~\cite{supp_Lieb}. A BEC at $\Gamma$ point becomes so unstable in Sp1 that it relaxes quickly into the true ground states. As for $|\chi|$, it strongly depends on the crossed FORT depths~\cite{supp_Lieb}. We note that the offset of $|\chi|$ in Sp2 data shifts upward since the regions of interest (blue and green circles in Fig.~\ref{image1}(e)) are close to $\Gamma$ points.

%
%

At a short ramp-up time of around 100~ms in Fig.~\ref{image2}(c), $C$ is positive, which means that the phase transition has already happened. At the same time, $|\chi|$ is still small, which indicates that the simultaneous occupation of both chiral modes is observed more often than symmetry-broken images are. In a previous study~\cite{Struck:2011}, the possibility of domain formation was mentioned; however, this has yet to be confirmed. Therefore, we conducted an investigation to clarify this issue. The direct observation of chiral-mode domains in optical lattices was proposed in~\cite{KosiorA2014} assuming the far-field regime, which is difficult to reach in our system since the trap frequency limits in-plane TOF. Instead, we attempt to detect the formation of the chiral domains by observing the interference of spatially separated chiral modes.

\begin{figure}[!tb]
\begin{center}
\includegraphics[width=80mm]{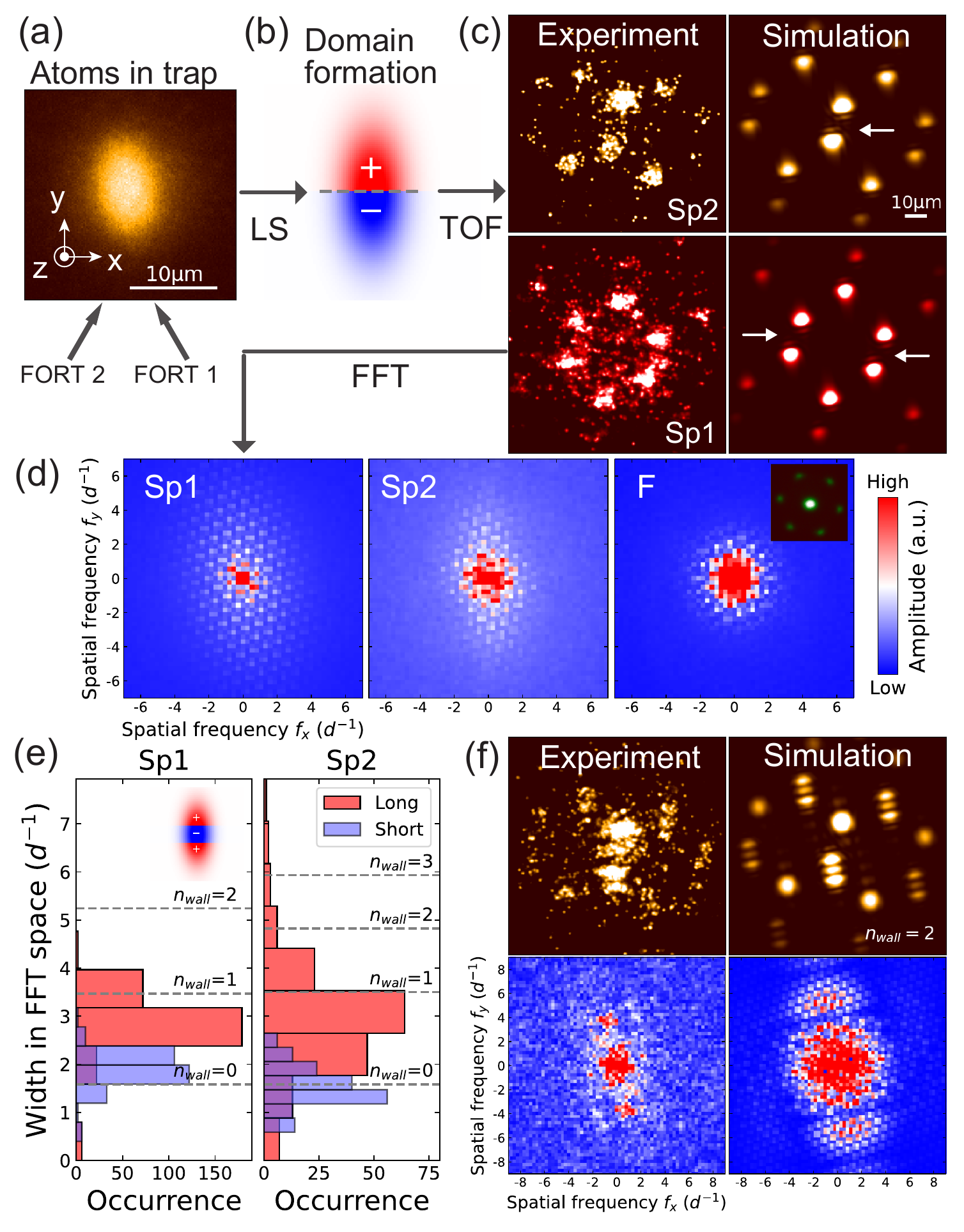}
\caption{Chiral-mode domains. (a) $In\mathchar`-situ$ image of atoms in the trap. The black arrows depict the directions along which the crossed FORTs are applied. The image is averaged over 10 runs. (b) Schematic of domain formation. The dotted line at the center is a linear domain wall. (c) Experimentally observed (left) and simulated (right) TOF images. The ramp-up time of 70~ms and 80~ms was used for Sp1 and Sp2, respectively. In the right figure, the white arrows point to the region where the fringes appear, and the color scale is saturated at 0.3 times the maximum probability to emphasize the fringes. (d) Amplitude spectra of FFT for Sp1, Sp2, and F. The color scale is saturated at 0.2 times the maximum amplitude. $d=\frac{2}{3}\lambda$ in the axis labels means the lattice constant. FFT images are averaged for more than 100 runs. The inset shows the averaged TOF image of F as an icon. The distance of the inherent interference peaks is 1.18~$d^{-1}$ in FFT space. (e) Histograms of the widths in FFT space. The red (blue) bars mean the widths along the long (short) axis. The dashed gray lines are the long-axis width estimated by the numerical simulation assuming the different number of walls $n_{wall}$. (f) TOF images that have multiple domain walls and their FFT signals. For better visibility, a Gaussian filter with $\sigma=0.53~d^{-1}$ is applied to the experimentally observed TOF images in (c) and (f). The maximum and offset of the color scale of the experimental data in (c) and (f) are 0.4 times and 0.04 times the maximum FL count, respectively. }
\label{image3}
\end{center}
\end{figure}

Figure~\ref{image3}(a-c) explains the processes from the domain formation to the observation of interference by the in-plane TOF. As shown in Fig.~\ref{image3}(a), the atoms in the optical triangular lattice and crossed FORTs have a shape elongated along the y-axis. We assumed that during the ramp-up time, a linear domain wall as shown in Fig.~\ref{image3}(b), which is the simplest of its kind, is formed most likely along the x-axis because the domain wall energy is proportional to its length. During in-plane TOF, wave packets with different chiral modes interfere with each other. Consequently, interference fringes are observed. Figure~\ref{image3}(c) shows single-shot images of Sp1 and Sp2 with fringes and simulated images. 
To visualize the effect of fringes, we applied a fast Fourier transform (FFT) to the TOF images. Figure~\ref{image3}(d) shows the amplitude spectrum of the FFT images averaged over more than 100 runs. The spectrum spreads over along the y-axis, which results from fringes. For comparison, the amplitude spectrum of F without lattice shaking is also shown. It appears symmetric, with no signs of fringes. We made a fit to the FFT signals to extract the widths along the long and short axes~\cite{supp_Lieb}. Figure~\ref{image3}(e) shows the histograms together with numerical simulation assuming various numbers of domain walls. While a single domain wall is dominant, multiple domain walls are sometimes formed in Sp2. Figure~\ref{image3}(f) shows the experimentally observed TOF image with multiple domain walls and the simulated image with $n_{wall}=2$. In the FFT signals, we can see side peaks that result from interference of the same chiral modes.

%
%

\begin{figure}[!tb]
\begin{center}
\includegraphics[width=80mm]{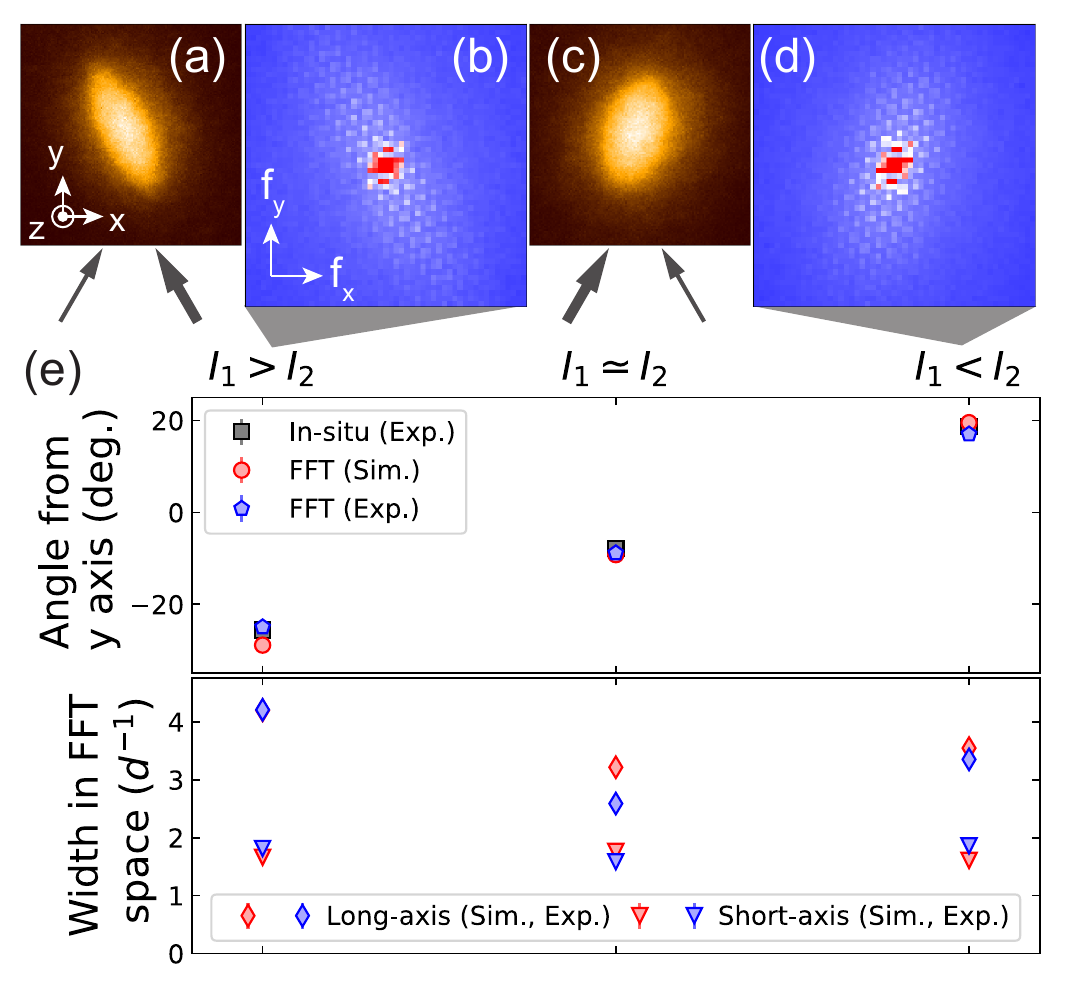}
\caption{Control of domain wall orientation. (a) Averaged $in\mathchar`-situ$ image of the atoms and (b) amplitude spectrum of FFT for TOF images in Sp2 in the condition $I_1>I_2$. (c),(d) Counterparts in the opposite condition $I_1<I_2$. (e) Angles in $in\mathchar`-situ$ images and FFT space (upper row), and widths in FFT space (lower row) at different balances of FORTs. Error bars denoting fitting errors are covered with markers.}
\label{image4}
\end{center}
\end{figure}

Finally, the orientation of the domain wall was controlled using crossed FORTs. The trap frequencies in the xy-plane are dominated by FORT~1 and FORT~2 (Fig.~\ref{image3}(a)), which have much tighter beam waists of 23 $\mu$m than the horizontal beam waists of the three beams that comprise the triangular lattice $\sim 120$ $\mu$m~\cite{RYamamoto:2020}. Therefore, the direction along which the atomic cloud is elongated depends on the intensity balance between FORT~1 and FORT~2. Figure~\ref{image4}(a) shows the atom distribution when the intensity of FORT~1, $I_{1}$, was much stronger than that of FORT2, $I_2$. The amplitude spectrum of the FFT in Sp2 is oriented in line with the atom distribution (Fig.~\ref{image4}(b)). Figure~\ref{image4}(c), (d) are the opposite case. Figure~\ref{image4}(e) compares the angles and widths in FFT space at different balances of FORTs. We can see good agreement between experiment and numerical simulation assuming $n_{wall}=1$.

%
%
In conclusion, we studied the relaxation from the ferromagnetic phase to two frustrated phases (Sp1 and Sp2) in the XY model on shaken optical triangular lattices. We revealed that domain walls~\cite{ParkerC2013,KockT2015,WangX2022} are formed in Sp1 and Sp2, which accounts for the simultaneous occupation of the two chiral modes. In this study, the system does not reach a strongly correlated regime. When the interaction is increased to $U/n|J| \gg 1$, one can access frustrated quantum magnetism, where the appearance of gapped spin-liquid phases is predicted~\cite{Eckardt:2010}. The system can be mapped onto the spin-1 quantum XY model~\cite{Altman:2002,DYamamoto:2020} near the Mott insulating state with unit filling. In this situation, the quantum phase transition between the chiral superfluid and Mott insulator, where the symmetry of U(1)$\times \mathbb{Z}_2$ is broken, can be investigated. Non-equilibrium dynamics after quenching across the phase transition and formation of domain structures can reveal the quantum Kibble-Zurek mechanism in this model~\cite{Keesling:2019}. 


%
%
\begin{acknowledgments}
We thank Daisuke Yamamoto for helpful discussions. This work was supported by JSPS KAKENHI Grant Numbers JP19H01854, 23H01133, ImPACT Programme of Council for Science, Technology and Innovation (Cabinet Office, Government of Japan), and JST ERATO-FS Grant Number JPMJER2204. 

\end{acknowledgments}

%
%
\clearpage

\begin{widetext}
\begin{center}
\textbf{\large Supplemental Material for \\ Observation of Chiral Domain Structures in a Frustrated XY Model on Optical Triangular Lattices}
\end{center}
\end{widetext}
\setcounter{equation}{0}
\setcounter{figure}{0}
\setcounter{table}{0}
\setcounter{page}{1}

\makeatletter
\renewcommand{\thesection}{S\arabic{section}}
\renewcommand{\theequation}{S\arabic{equation}}
\renewcommand{\thefigure}{S\arabic{figure}}
\makeatother

\onecolumngrid

%
%
\section{Phase diagram of a frustrated XY model}

In a tight-binding approximation, the renormalized dispersion relation in the periodically driven triangular lattice is
\begin{equation}
\epsilon(J,J',\mathbf{k}) = -2J {\rm cos}(k_y d) - 4J'{\rm cos}\left(k_xd\sqrt{3}/2 \right){\rm cos}\left( k_yd/2 \right),
\label{eq:effectiveBandEnergy}
\end{equation}
where $d= \frac{2}{3}\lambda$ is the lattice constant. The parabolic dispersion relation term along the direction perpendicular to the triangular lattice plane is omitted.
According to the dispersion relation Eq.~\ref{eq:effectiveBandEnergy}, the minima of energy $\epsilon(J,J',\mathbf{k})$ are given by
\begin{eqnarray}
    \mathbf{q}_{gs}(J,J')= \left\{
    \begin{array}{llll}
    (0,0) & {\rm for } \; J'> & -2J \; & {\rm and } \; J'>0 \;{\rm (F)} \\
    \frac{2}{d}\left( \frac{\pi}{\sqrt{3}}, 0 \right) &  {\rm for } \; J'< & 2J \; & {\rm and } \; J'<0 \;{\rm (R)}  \\
    \frac{2}{d}\left(\frac{\pi}{\sqrt{3}}, \pm {\rm arccos} \left( \frac{J'}{2J} \right) \right) & {\rm for } \; J'> & 2J \; & {\rm and } \; J'<0 \;{\rm (Sp1)} \\
    \frac{2}{d}\left(0, \pm {\rm arccos} \left( -\frac{J'}{2J} \right) \right) & {\rm for } \; J'< & -2J \; &{\rm and } \; J'>0 \;{\rm (Sp2)} ,
    \end{array}  
    \right.
    \label{eq:minimaOfXYmodel}
\end{eqnarray}
where we assigned the same names as in~\cite{Struck:2011} to the spin configurations that are realized with the different spin-spin couplings $J, J'$. The corresponding phase $\theta_i^{gs}$ at each lattice site $i$ is given by:
\begin{equation}
    \theta_i^{gs} = \mathbf{q}_{gs} \cdot \mathbf{R}_i,
\end{equation}
where $\mathbf{R}_i$ is a lattice vector. Equation \ref{eq:minimaOfXYmodel} does not include the cases for $J'=0$ where the system forms decoupled chains and the dispersion relation along $k_x$ becomes degenerate. Energy of dispersion relation at the position of minima $\epsilon(J, J',\mathbf{q}_{gs})$ is equivalent to the ground state energy of the classical XY model per particle $E_{XY}^{gs}(J, J')/N$:
\begin{eqnarray}
    \epsilon(J,J',\mathbf{q}_{gs})=\frac{E_{XY}^{gs}(J,J')}{N} = \left\{
    \begin{array}{ll}
    -J-2|J'| & {\rm for} \; |J'|>-2J \\
    J+\frac{J'^2}{2J} & {\rm for} \; |J'|<-2J.
    \end{array}
    \right.
    \label{eq:phaseDiagramOfXYmodel}
\end{eqnarray}

\begin{figure}[tb]
\begin{center}
\includegraphics[width=160mm]{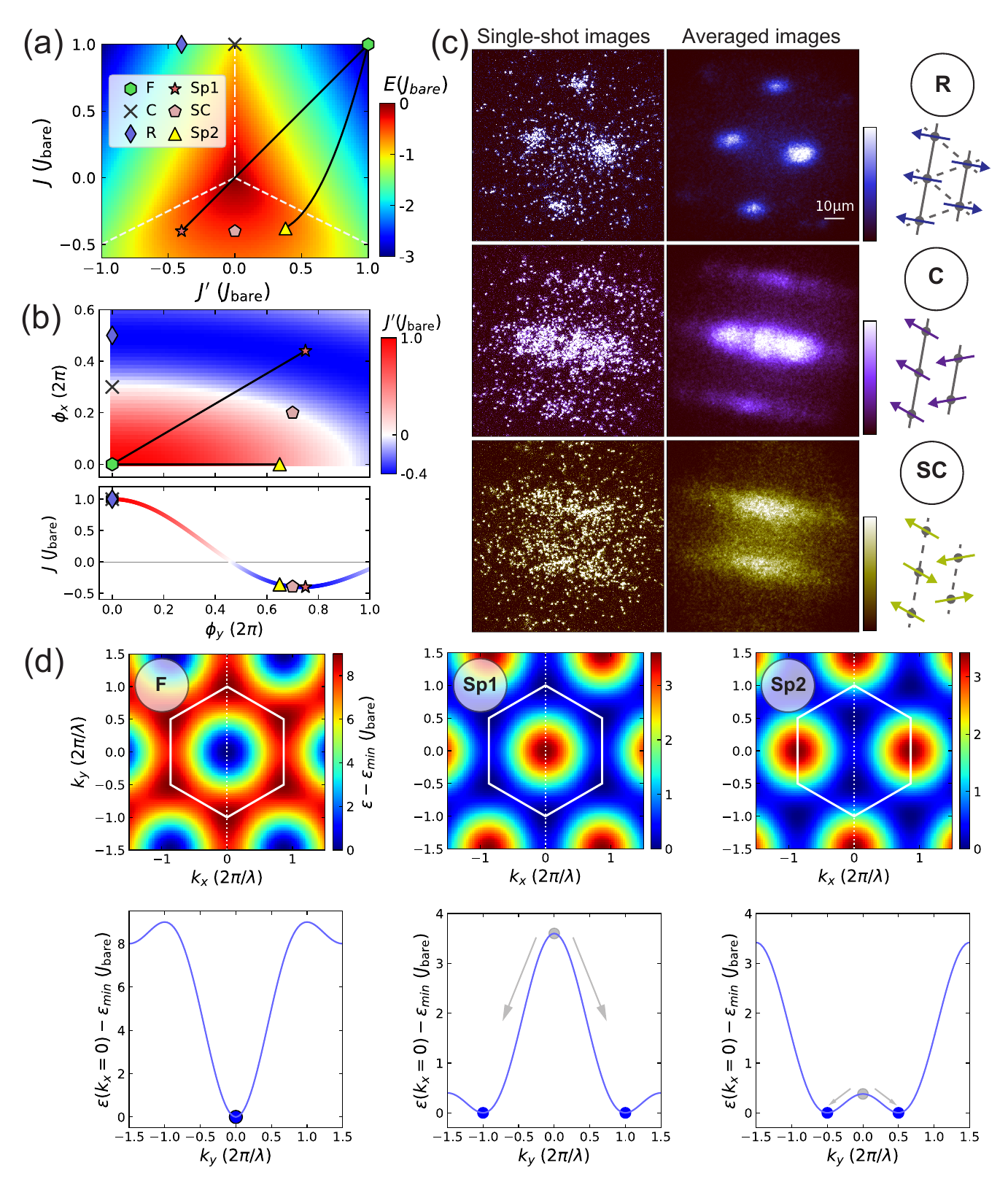}
\caption{(a) Phase diagram of the classical XY model in a triangular lattice. The tunneling parameters at each phase are $(J,J')/J_{\rm bare}=(1,1)$ for Ferro~(F), $(1,-0.4)$ for Rhombic~(R), $(1,0)$ for 1D chains~(C), $(-0.4,0)$ for Staggered 1D chains~(SC), $(-0.35,-0.35)$ for Spiral~1~(Sp1), and $(-0.35,0.35)$ for Spiral~2~(Sp2). These parameters are used in Fig.~1(c),(d) of the main text, and Fig.~\ref{figureS1}(c). The color scale represents the energy in Eq.~\ref{eq:phaseDiagramOfXYmodel}. (b) $J,J'$ in Eqs.~\ref{eq:J},~\ref{eq:Jp} as a function of $\phi_x,\phi_y$. The black solid lines in (a),(b) represent the paths from F to Sp1 or Sp2. (c) Interference patterns of R, C, and SC. The arrows on the triangular lattice mean the spin states. For the averaged images, 20 independent experimental realizations were used. (d) Band dispersion relations of F (left), Sp1 (middle), and Sp2 (right) based on Eq.~\ref{eq:effectiveBandEnergy}. The white hexagonal lines in the upper row mean the 1st Brillouin zone. The lower row shows the cross section along the white dotted lines in the upper row. The blue filled circles denote the quasimomenta of the ground states in Eq.~\ref{eq:minimaOfXYmodel}. During the Floquet drive, a BEC at $\Gamma$ point in Sp1 and Sp2, expressed as the gray transparent circles, relaxes toward the ground states. }
\label{figureS1}
\end{center}
\end{figure}

Figure~\ref{figureS1}(a) shows a zero-temperature phase diagram of the classical XY model in a triangular lattice based on Eq.~\ref{eq:phaseDiagramOfXYmodel}.

The expressions of the phase modulation signals for the triangular lattice beams can be written as

\begin{eqnarray}
\delta \phi_1 (t) &=&  \phi_x {\rm {sin}}(\Omega t) + \phi_y {\rm{cos}}(\Omega t) \nonumber \\
&=& \sqrt{\phi_x^2 +\phi_y^2} {\rm{sin}}(\Omega t + \alpha) \\
\delta \phi_2 (t) &=& -\phi_x {\rm {sin}}(\Omega t) + \phi_y {\rm{cos}}(\Omega t) \nonumber \\
&=&\sqrt{\phi_x^2 +\phi_y^2} {\rm{sin}}(\Omega t + \pi - \alpha)  \\
\alpha &=&{\rm arcsin}\left(\frac{\phi_y}{\sqrt{\phi_x^2+\phi_y^2}} \right),
\end{eqnarray}
where $\Omega/2\pi$ is the driving frequency, and $\phi_{x},\phi_{y}$ are the modulation amplitudes.
Thus, the effective tunneling processes are expressed as
\begin{eqnarray}
J &=& \mathcal{J}_0(F_A)J_{\rm{bare}} \label{eq:J}\\
J' &=& \mathcal{J}_0(F^{\prime}_A)J_{\rm{bare}} \label{eq:Jp}\\
F_A &=& \frac{m \Omega d^2}{\hbar} \frac{\phi_y}{2 \pi} \label{eq:K}\\
F^{\prime}_A &=& \frac{m \Omega d^2}{2\hbar} \frac{\sqrt{9 \phi_x^2 + \phi_y^2}}{2 \pi}, \label{eq:Kp}
\end{eqnarray}
where $\mathcal{J}_0$ is the zeroth order Bessel function of the first kind. Figure~\ref{figureS1}(b) plots $J,J'$ as a function of $\phi_x, \phi_y$. Figure~\ref{figureS1}(c) shows experimentally observed interference patterns of unfrustrated phases.
Figure~\ref{figureS1}(d) shows the effective band dispersion relations of F, Sp1, and Sp2 based on Eq.~\ref{eq:effectiveBandEnergy} using $J, J'$ in Fig.~\ref{figureS1}(a),(b).

%
%
\section{Experimental sequence and micromotion effect}

\begin{figure}[tb]
\begin{center}
\includegraphics[width=180mm]{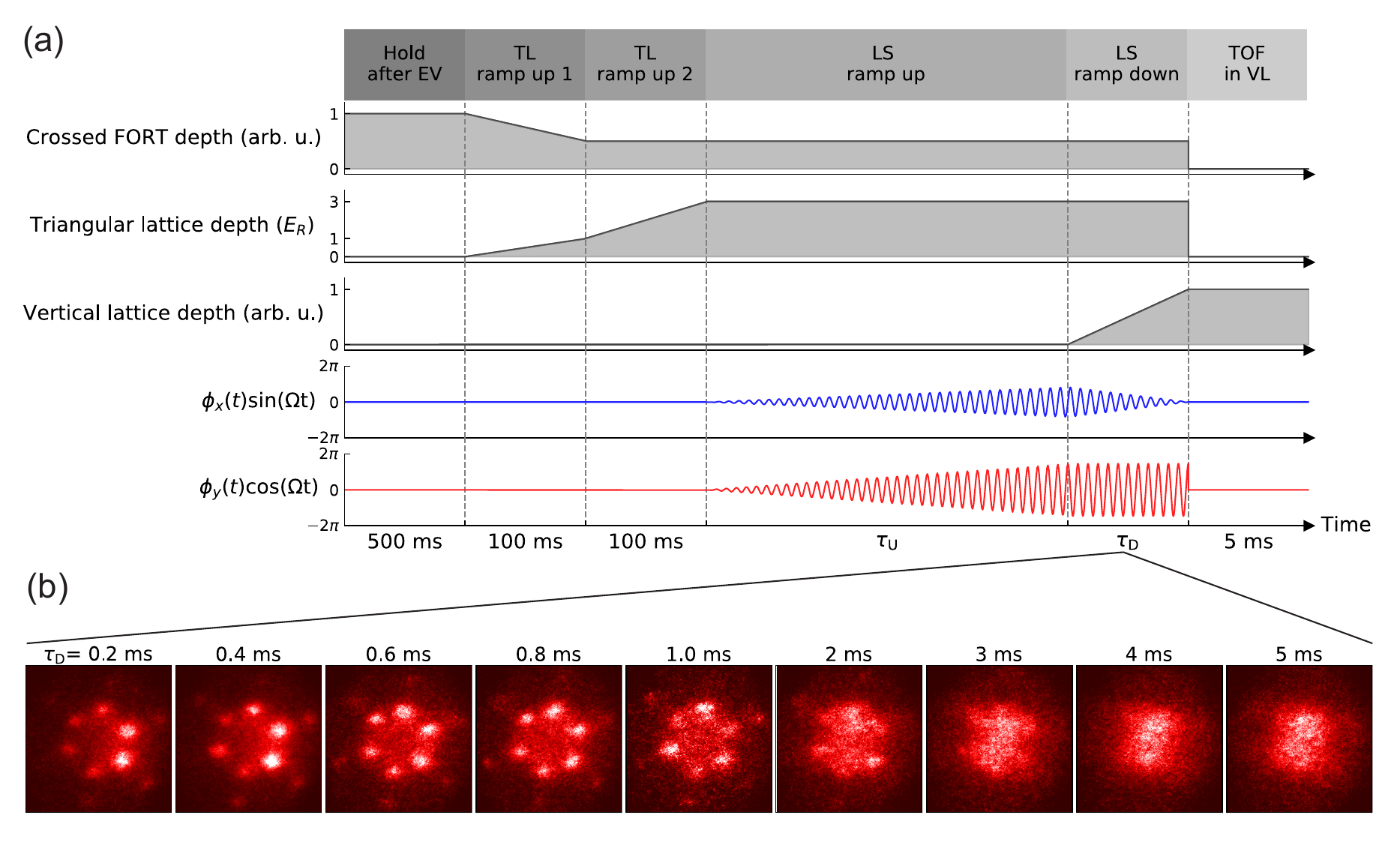}
\caption{(a) Experimental sequence before, during, and after the Floquet drive. The horizontal axis is not scaled by the times at each column. The blue and red lines mean the phase modulation signals. The abbreviations EV, TL, LS, and VL mean evaporative cooling, triangular lattice, lattice shaking, and vertical lattice, respectively. (b) In-plane TOF images with various ramp-down times $\tau_{\rm D}$. Each image is averaged for more than 20 runs. A too short $\tau_{\rm D}$ results in the shift of the envelope centers on account of the micromotion; on the other hand, a too long $\tau_{\rm D}$ causes relaxation from Sp1 to Sp2 via Staggered 1D chains.}
\label{figureS2}
\end{center}
\end{figure}

Figure~\ref{figureS2} shows the experimental sequence from the holding time of the crossed FORTs after evaporative cooling to in-plane TOF. After 500 ms of holding atoms in the crossed FORTs for thermalization, we ramped up the optical triangular lattice to $V_0=1.0~E_R$ in 100~ms and simultaneously ramped down the crossed FORTs to a moderate depth so that the Flquet evaporation could efficiently work (see the S.4 Flquet evaporation). In the 2nd ramp of 100~ms, the optical triangular lattice was increased to $V_0=3.0~E_R$. Subsequently, the phase modulation amplitudes to drive the positions of the optical triangular lattice were ramped up with various times $\tau_{\rm U}$ as in Fig.~2 of the main text. Before the in-plane TOF, the vertical lattice depth increased to $V_{\rm ver}/k_B=285$ nK. Simultaneously, one of the phase modulation signals was decreased to zero to eliminate the micromotion effect that inherently arises in Floquet systems~\cite{Goldman:2014, RevModPhys.89.011004, Guo:2019}. Figure~\ref{figureS2}(b) displays the in-plane TOF images in Sp1 with several ramp-down times, $\tau_{\rm D}$. In the case of $\tau_{\rm D} < 2\pi/\Omega =0.833$ ms, the atoms undergo a momentum kick (i.e., a micromotion) at the end of elliptical modulation of the lattice position. Consequently, the centers of the envelope in the momentum distributions shift. In contrast, if $\tau_D\ge 2\pi/\Omega$, the elliptical modulation reduces to a liner modulation, where the micromotion can be removed as long as the final phase of the remaining modulation signal is set to fulfill arg$[\phi_y(\tau_{\rm U}+\tau_{\rm D}) {\rm cos}(\Omega (\tau_{\rm U}+\tau_{\rm D}))] = \pm \pi/2$. A too long $\tau_{\rm D}$, however, leads to relaxation from Sp1 to Sp2 via Staggered 1D chains (see Fig.~\ref{figureS1}(b)). We set $\tau_{\rm D}=1.0$ ms in our experiments.

%
%
\section{Multi-photon interband excitation in a driven triangular lattice}

In quantum systems with Flquet engineering, the heating effect during the periodic drive is the most serious problem. In the case of driven optical lattices, the major cause of this heating is multi-photon excitation between time-averaged energy bands~\cite{Arlinghaus:2010, Weinberg:2015}. The periodic modulation with frequency $\Omega/2\pi$ causes the interband coupling processes that conserve quasimomentum but change the energy by integer multiples of the photon energy $\hbar \Omega$. An $n_{p}^{\rm th}$-order multi-photon transition is assumed to occur when the following resonance condition is fulfilled:
\begin{equation}
n_p \times \hbar \Omega = \Delta E_{\beta}^{\rm eff} (\mathbf{q}=\mathbf{q}_{gs}),
\label{eq:MPA}
\end{equation}
where $\Delta E_{\beta}^{\rm eff}(\mathbf{q}) = \epsilon_{\beta}^{\rm eff}(\mathbf{q}) - \epsilon_{0}^{\rm eff}(\mathbf{q})$, and $\epsilon_{\beta}^{\rm eff}(\mathbf{q})$ is the time-averaged single-particle energy of $\beta$-th band with quasimomenta $\mathbf{q}$.

\begin{figure}[tb]
\begin{center}
\includegraphics[width=160mm]{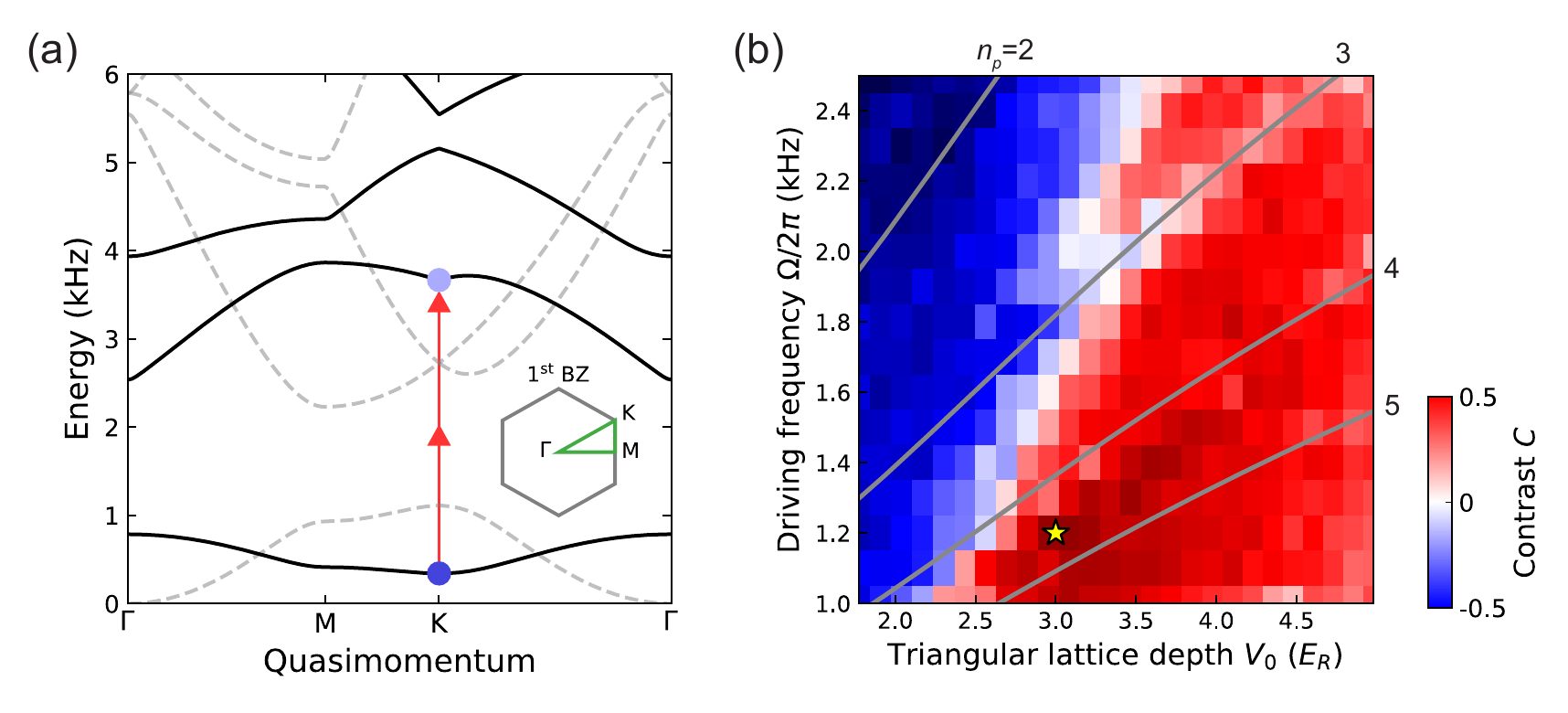}
\caption{Multi-photon interband transition in a driven triangular lattice. (a) Illustration of two-photon transition from the ground to the first excited band at $K$ point for Sp1. The gray dashed lines mean the bare single-particle band energy with the triangular lattice depth $V_0=1~E_R$. The black solid lines plot the time-averaged band energy with $(J,J')/J_{\rm bare} = (-0.4,-0.4)$. (b) Excitation spectrum for Sp1. The gray solid lines plot the resonance conditions in Eq.~\ref{eq:MPA} from the lowest to the first excited band at $K$ point with different $n_p$. The yellow star means the parameters we use in the main text.}
\label{figureS3}
\end{center}
\end{figure}

Figure~\ref{figureS3}(a) shows the effective energy bands of the triangular lattice with $V_0=1.0~E_R$ in the case of Sp1. The red arrows indicate the two-photon transition from the lowest to the first excited band. These interband transitions caused by periodic driving significantly reduce the contrast $C$ defined in Fig.~2(a). The decrease in $C$ can be attributed to two distinct processes. First, the atoms excited into higher bands are no longer trapped in the optical lattice. Thus, the atomic losses reduce the bosonic enhancement $n|J^{(\prime)}|$. Second, an interacting BEC in the excited bands rapidly decays owing to the scattering processes, which results in a decrease in the degree of coherence in the system~\cite{Martikainen:2011, Paul:2013}.

Figure~\ref{figureS3}(b) shows the results of a spectroscopic study of these multi-photon transitions in the driven triangular lattice. The excitation spectra were taken with various lattice depths at fixed forcing amplitudes $F_A,F^{\prime}_A$. The resonant driving frequencies in Eq.~\ref{eq:MPA} with different $n_p$ are plotted together with the excitation spectra. These $ab \; initio$ calculations agree with the experimental data, and multi-photon excitations up to the fourth-order are visible, which is consistent with a previous study~\cite{Weinberg:2015}. We note that to keep $F_A,F^{\prime}_A$ constant throughout the spectroscopy, the final amplitudes of phase modulation $\phi_x(\tau_{\rm U}),\phi_y(\tau_{\rm U})$ were adjusted according to Eqs.~\ref{eq:K},~\ref{eq:Kp}. The maximum $C$ was reached within a narrow parameter region between $n_p=4$ and $n_p=5$. In the main text, we use $V_0=3.0~E_R$ and $\Omega/2\pi=1.2$ kHz for this reason.

%
%
\section{Floquet evaporation}

\begin{figure}[tb]
\begin{center}
\includegraphics[width=180mm]{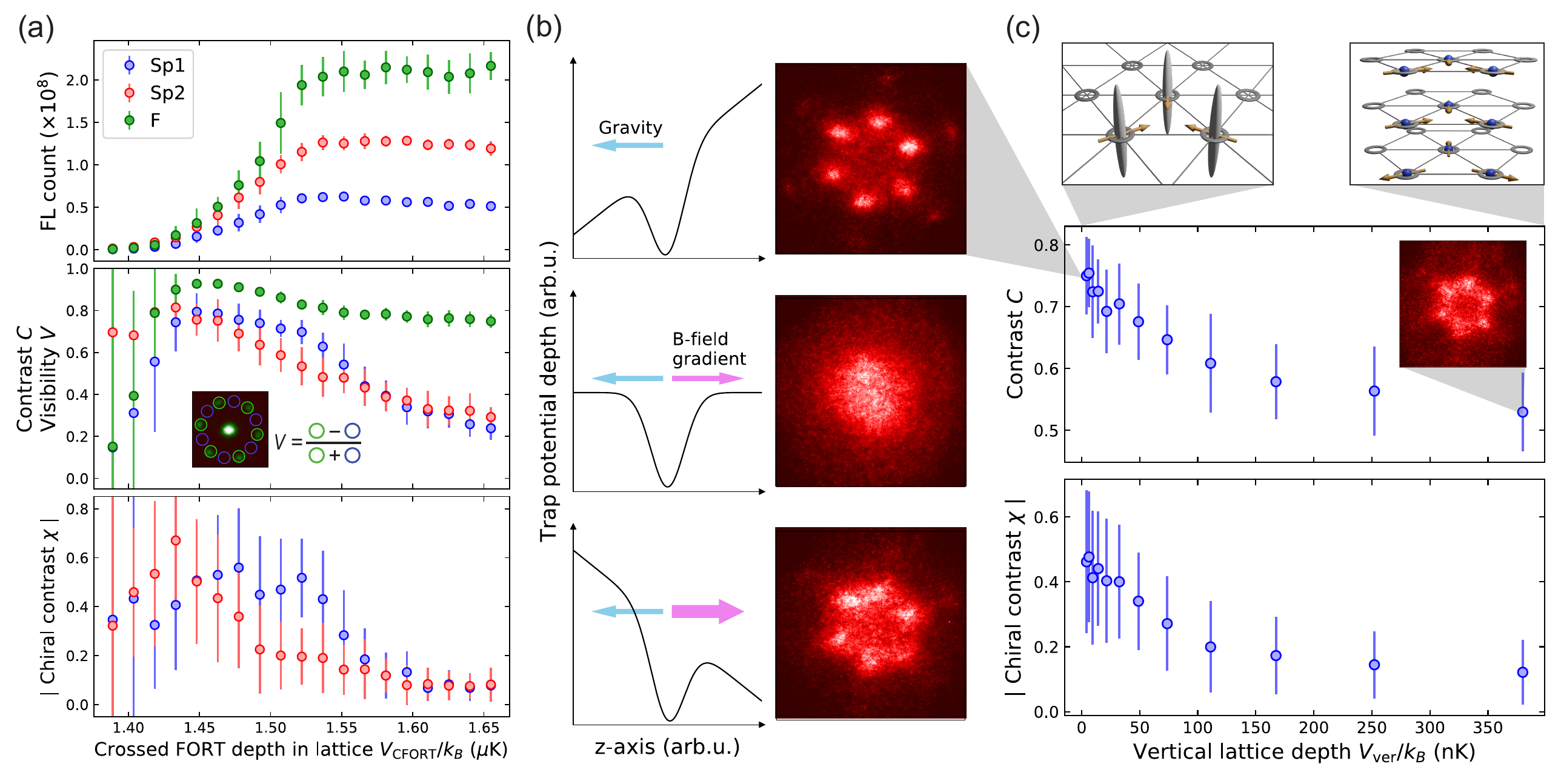}
\caption{Floquet evaporation. (a) Dependence on the crossed FORT depth $V_{\rm CFORT}$ during the Floquet drive. The LS ramp-up time~$\tau_{\rm U}$ is fixed at 300 ms. Error bars denote standard deviations (SD). The inset in the middle shows an interference pattern of F and the definition of visibility $V$. The error bars in $C$, $V$, and $|\chi|$ become large in the shallow $V_{\rm CFORT}$ as the number of atoms decreases. (b) Dependence on the B-field gradient along the z-axis. The black solid lines depict the external potential with gravity and B-field gradient. The column on the right shows the corresponding TOF images. (c) Dependence on vertical lattice depth $V_{\rm ver}$ during Floquet drive.}
\label{figureS4}
\end{center}
\end{figure}

As explained in the previous section, the triangular lattice depth $V_0$ and driving frequency $\Omega/2\pi$ are important parameters to avoid multi-photon interband transitions. Another salient factor that should be considered to reduce heating in Floquet-engineered quantum gases is Floquet evaporation~\cite{PhysRevLett.119.200402}. As in Fig.~\ref{figureS4}(a), we found that the crossed FORT depth during the Floquet drive should be decreased such that the resonantly scattered atoms can leave the trap before dissipating their energy into the system. We examined the effect of Floquet evaporation by applying a B-field gradient along the z-axis (Fig.~\ref{figureS4}(b)). Without the B-field gradient, heated atoms can evaporate because of gravity. When a B-field gradient was applied to cancel the gravity, the atomic ensemble suffered from heating and exhibited no interference peaks. When the B-field gradient was stronger than gravity, the visibility recovered, as expected.

Figure~\ref{figureS4}(c) shows the dependence of $C$ and $|\chi|$ for Sp1 on the vertical lattice depth $V_{\rm ver}$ in the driven triangular lattice. As $V_{\rm ver}$ increases, both $C$ and  $|\chi|$ decrease, which indicates that Floquet evaporation along the z-axis does not work well in the presence of the vertical lattice because it prevents thermal atoms from evaporating along the gravitational direction. This could be a crucial problem for the experimental realization of the strong interaction regime ($U/n|J| \gg 1$) in the two-dimensional driven lattice systems.

\begin{figure}[tb]
\begin{center}
\includegraphics[width=180mm]{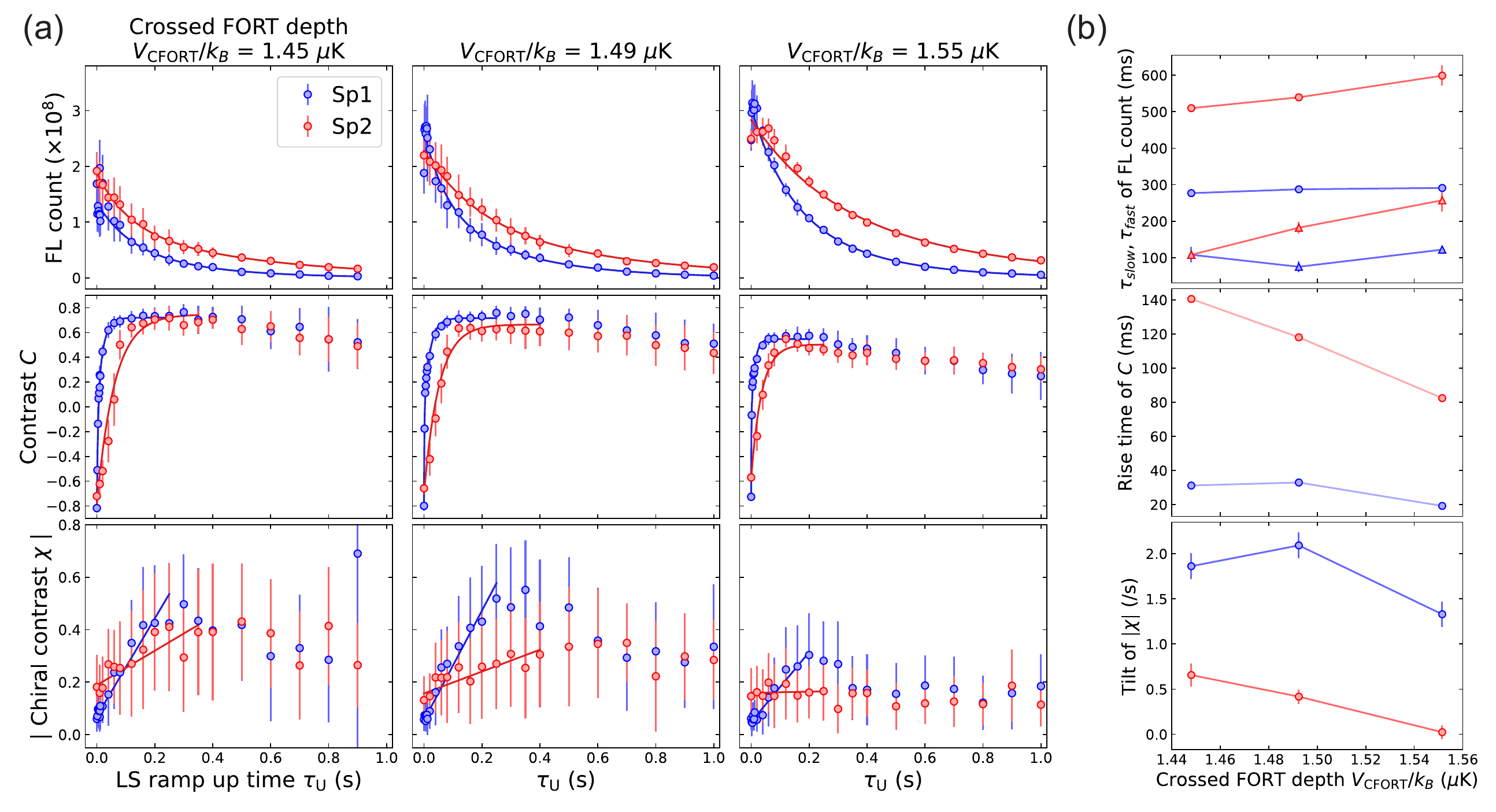}
\caption{Dependence on crossed FORT depth~$V_{\rm CFORT}$ and LS ramp-up time~$\tau_{\rm U}$. (a) FL count, $C$, and $|\chi|$ with various ramping times at three different depths of crossed FORT. The solid lines are fitting results. The fitting functions for FL count are a double exponential function with two time scales~$(\tau_{\rm slow}, \tau_{\rm fast})$. Error bars mean SD. (b) The decay of FL count, the rise time of $C$, and the tilt of $|\chi|$ at each crossed FORT depth in (a). These values are extracted from the fitting results. Error bars denote fitting errors.}
\label{figureS5}
\end{center}
\end{figure}
We systematically investigated dependence on the crossed FORT depth and LS ramp-up time. Figure~\ref{figureS5}(a) shows the results. The contrast and chiral contrast of the middle column in Fig.~\ref{figureS5}(a) correspond to Fig.~2(c) in the main text. Figure~\ref{figureS5}(b) summarizes the fitting results. These results and Fig.~\ref{figureS4}(a) ensure that although the crossed FORT depth affects the maximum values of $C$ and $|\chi|$, it does not change the fact that the relaxation from F to Sp1 is much faster than that from F to Sp2.

FL count and $C$ in Fig.~\ref{figureS5}(a) are fitted by single or double exponential functions, while $|\chi|$ is fitted by a linear function with offset. Ideally, $|\chi|$ might also be fitted by an exponential-like function in the same way as the contrast $C$. In reality, however, the loss and heating of atoms cause both $C$ and $|\chi|$ to decrease in long ramp-up times. Since these loss and heating mechanisms are too complex, we excluded the experimental data of the long ramp-up times from fitting. Unlike $C$, the plateau of $|\chi|$, which is important if one wants to make a fit to the data with an exponential-like function, can hardly be seen. Therefore, we approximated the fitting function to a linear one in the case of $|\chi|$.

%
%
\section{Relaxation from Rhombic to Spiral~1 and Spiral~2}

\begin{figure}[tb]
\begin{center}
\includegraphics[width=180mm]{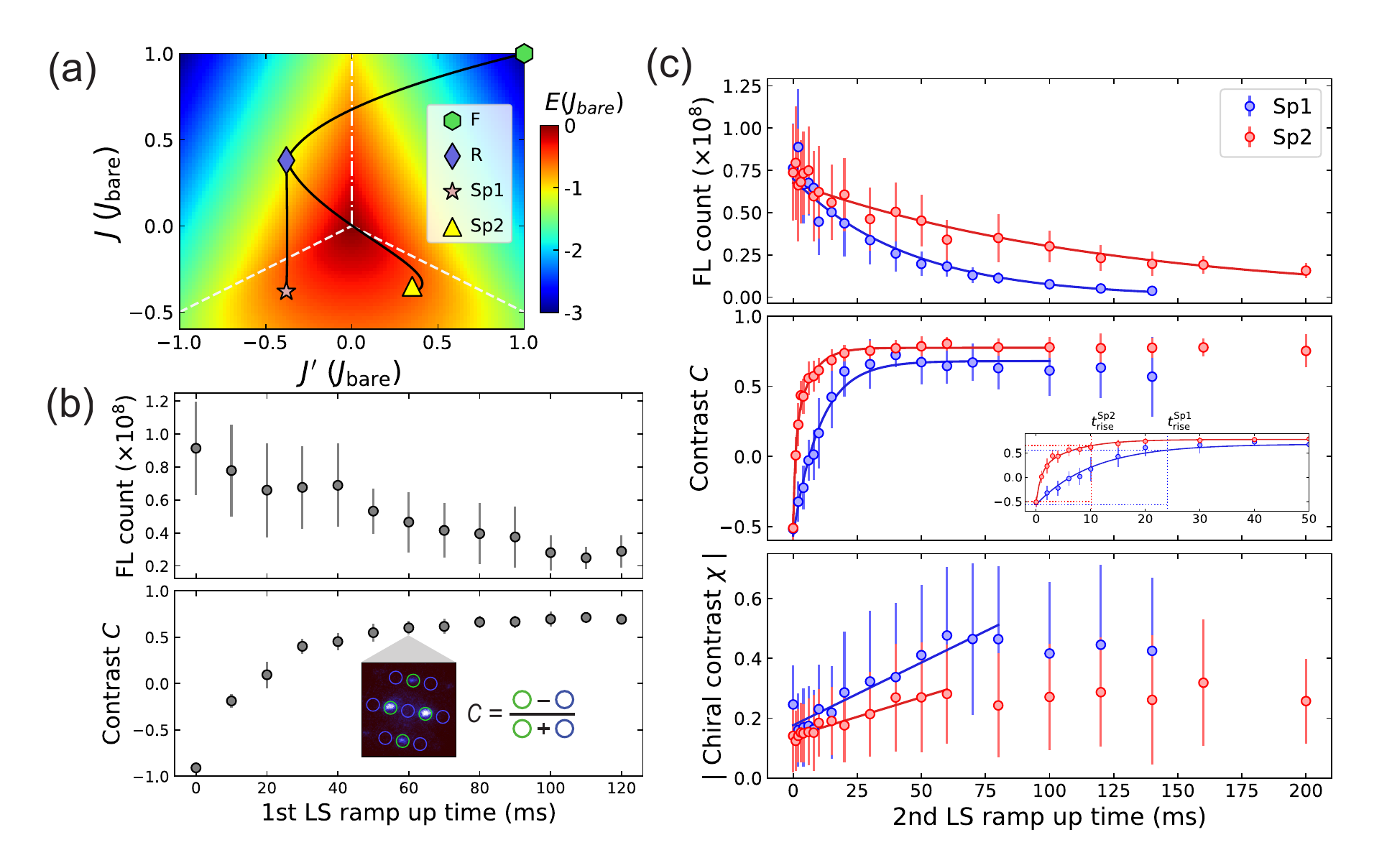}
\caption{Relaxation from R to Sp1 and Sp2. (a) The black lines on the phase diagram show the paths from F to Sp1 and Sp2 via R. (b) Relaxation from F to R. The inset shows the interference pattern at the ramp-up time of 60 ms and the definition of the contrast. Error bars mean SD. (c) Relaxation from R to Sp1 and Sp2. The solid lines are fitting results to the data.}
\label{figureS6}
\end{center}
\end{figure}

In Fig.~2(c) of the main text, we observed relaxation from F to Sp1 and Sp2. We found that the rise time of contrast $C$ differs between Sp1 and Sp2, which we attribute to the length of the path and the effective band structure. To ensure this scenario, we checked relaxation from R to Sp1 and Sp2 as seen in Fig.~\ref{figureS6}(a), which is the opposite case of relaxation from F to Sp1 and Sp2 used in the main text. Therefore, the relaxation time to Sp2 should be faster than that to Sp1.

We used a two-step ramp-up of the Floquet drive. The first ramp-up is to prepare R as an initial state. Figure~\ref{figureS6}(b) shows relaxation from F to R with various ramp-up times. We set the first ramp-up time at 60 ms, where atoms already become an R state. In the second ramp-up, atoms relax from R to Sp1 and Sp2. Figure~\ref{figureS6}(c) shows the results. As expected, the rise time of $C$ in Sp1 $t_{\rm rise}^{\rm Sp1} = 24.0$ ms is slower than that in Sp2 $t_{\rm rise}^{\rm Sp2} = 10.0$ ms, which is the opposite case of Fig.~2(c) in the main text. In making a fit to $C$, the definition of the empirical function in Eq.~2 is reversed such as

\begin{eqnarray}
f(t) =  \left\{ 
    \begin{array}{ll}
    A e^{-t/\tau} + B & {\rm for} \; {\rm Sp1}, \\
    \frac{A}{2} \left( e^{-t/\tau_{\rm fast}} + e^{-t/\tau_{\rm slow}} \right) + B & {\rm for} \; {\rm Sp2}. \\
    \end{array}
\right.
\label{eq:FitFuncC2}
\end{eqnarray}

The reason why a double-exponential function is used for Sp1 in Eq. 2 and for Sp2 in Eq.~\ref{eq:FitFuncC2} is because atoms go through a non-condensate state around the center of the phase diagram. There are two distinct processes; First, the distribution of atoms swiftly spreads in momentum space. Second, the atoms condensate again at a true ground state relatively slowly. FL count in the second ramp-up is fitted by a single exponential function instead of a double exponential function because the initial fast decay has already occurred in the first ramp-up. $|\chi|$ is fitted by a linear function with offset, the same as in the main text.

%
%
\section{Details of fitting to FFT signals} \label{sec6}
In Fig.~3(e) and Fig.~4(e) of the main text, the angle in $in\mathchar`-situ$ images and FFT images and the widths in FFT images are plotted. In this section, we describe how the information was obtained.

\begin{figure}[tb]
\begin{center}
\includegraphics[width=180mm]{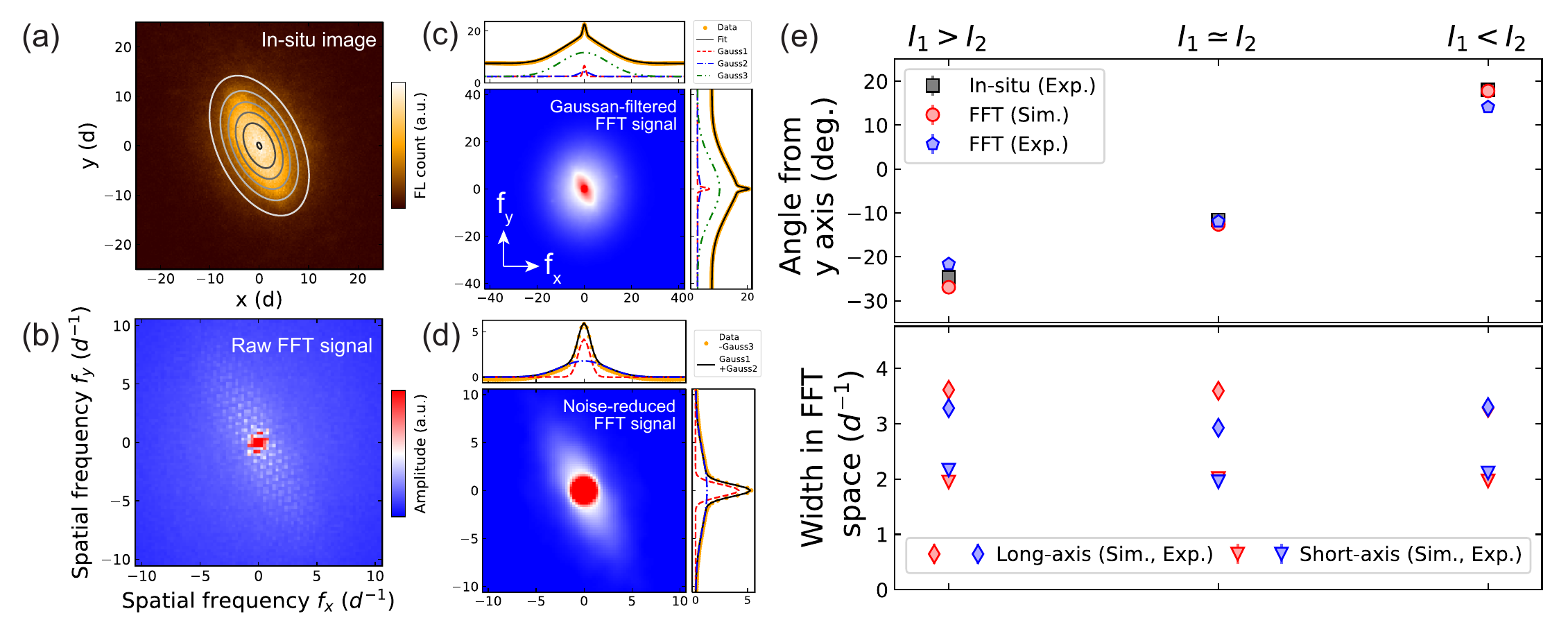}
\caption{Details of fitting. (a) A two-dimensional Gaussian fit with tilt angle was made to the $in\mathchar`-situ$ images used in Fig.~4(a). The gray contours represent the fitting results. (b) Raw FFT signal of Sp2 used in Fig.~4(b). (c) Gaussian-filtered FFT signal. The top and right sub-plots show the experimental data (orange circles) and fitting results (black solid line) accumulated along the x-axis and y-axis, respectively. Red, blue, and green lines mean Gauss1, Gauss2, and Gauss3 in Eq.~\ref{eq:TriGauss}, respectively. (d) FFT signal with Gauss3 subtracted. The color scale is saturated at the maximum of Gauss2. (e) Sp1 version of Fig.~4(e) in the main text.}
\label{figureS7}
\end{center}
\end{figure}

As in Fig.~\ref{figureS7}, the $in\mathchar`-situ$ images were fitted by a two-dimensional~(2D) elliptic Gaussian function with tilt and offset to know the angles and widths of the atomic distribution before in-plane TOF. The fitting results were used for the numerical simulation of spin domains, the details of which are described in the following section~\ref{sec7}.

As for the FFT signals, we extracted the information about angles and widths by using the following triple 2D Gaussian function;
\begin{eqnarray}
f_{FFT}(f_x,f_y) &=& A_1 \times {\rm exp}\left[ -\frac{(f_x {\rm cos}\theta+f_y {\rm sin} \theta)^2 }{2 \sigma_0^2} -\frac{(f_x {\rm sin}\theta-f_y {\rm cos} \theta)^2 }{2 \sigma_1^2} \right] \nonumber \\
&& + A_2 \times {\rm exp}\left[ -\frac{(f_x {\rm cos}\theta+f_y {\rm sin} \theta)^2 }{2 \sigma_2^2} -\frac{(f_x {\rm sin}\theta-f_y {\rm cos} \theta)^2 }{2 \sigma_3^2} \right] \nonumber \\
&& + A_3 \times {\rm exp}\left[ -\frac{f_x^2+f_y^2}{2\sigma_4^2} \right] + B,
\label{eq:TriGauss}
\end{eqnarray}
where $A_1,A_2,A_3, \sigma_0, \sigma_1, \sigma_2, \sigma_3, \sigma_4, \theta, B$ are fitting parameters. The first 2D Gaussian~(Gauss1) is for the inherent interference pattern of the triangular lattice, the second 2D Gaussian~(Gauss2) for the fringes originating from spin domains, and the third 2D Gaussian~(Gauss3) for noises in the experimental data. Gauss1 and Gauss2 have the same form. Because of this, although different initial guesses were used for $(\sigma_0,\sigma_1)$ and $(\sigma_2,\sigma_3)$ in making a fit, they were sometimes swapped. So, they were sorted after the fit. Before the fit, we applied a Gaussian filter with $\sigma=0.53~d^{-1}(=2~{\rm pixels})$ to the raw FFT signals, as in Fig.~\ref{figureS7}(b), to smooth out the spikes of the inherent interference patterns, whose spacing in FFT signal is $1.18~d^{-1}$. Figure~\ref{figureS7}(c) shows the Gaussian-filtered FFT signal of Sp2 together with the fitting results accumulated along the x-axis or y-axis. Figure~\ref{figureS7}(d) shows the FFT signals with Gauss3 subtracted to clarify that the experimental data is well fitted by Gauss1 and Gauss2.

Figure~4 in the main text shows the angle $\theta$ and widths $(\sigma_2,\sigma_3)$ of Gauss2 in Eq.~\ref{eq:TriGauss} using the data of Sp2. We conducted the same analysis on the data of Sp1 and confirmed that the experimental data agree with the numerical simulation, as in Fig.~\ref{figureS7}(e).

%
%
\section{Simulation of spin domains} \label{sec7}
\begin{figure}[tb]
\begin{center}
\includegraphics[width=180mm]{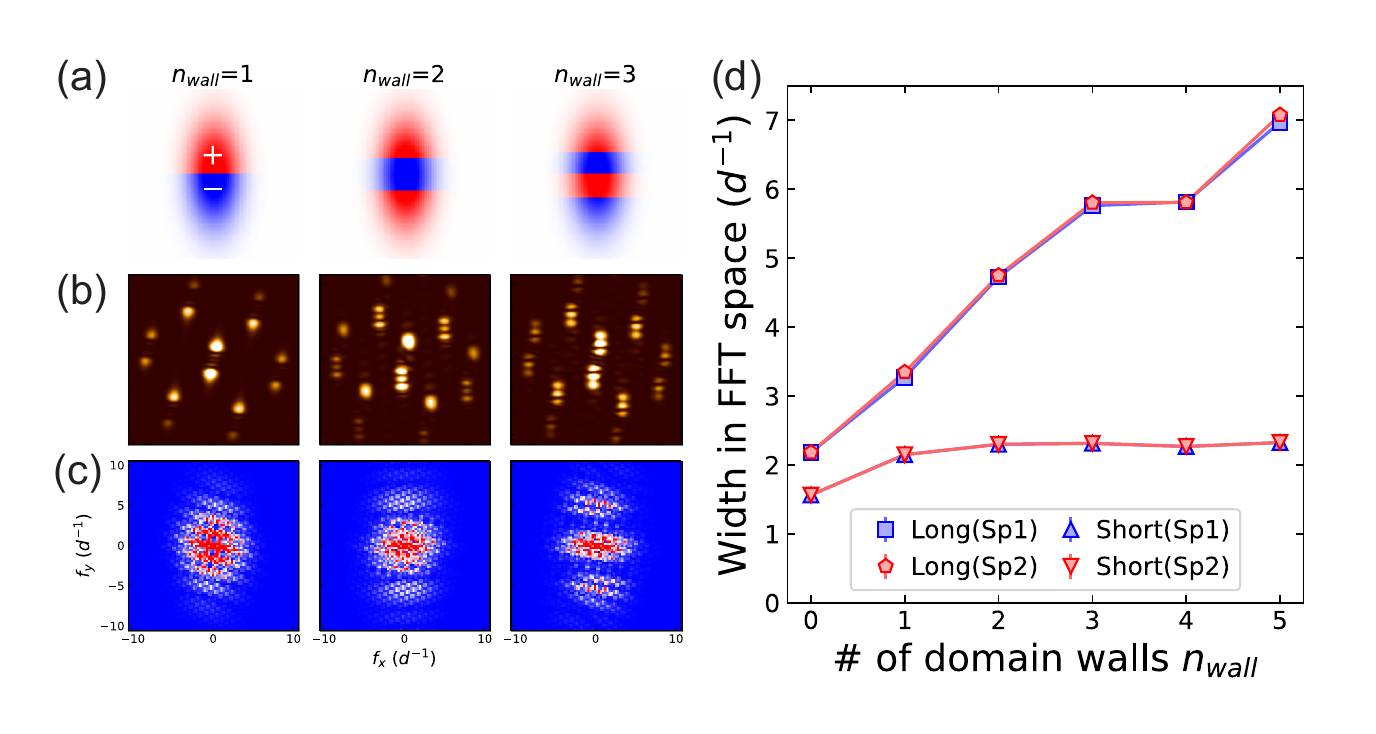}
\caption{Simulation of spin domains. (a) Schematics of the initial distribution of atoms with different numbers of domain walls. (b) Numerical simulation of TOF images in Sp2 at each number of domain walls. The color scale is saturated at 0.3 times the maximum value to highlight the fringes. (c) FFT signals of the TOF images in (b). The color scale is saturated at 0.2 times the maximum value to emphasize the high-frequency components. (d) Widths along the long and short axes in FFT space at each number of domain walls. Error bars denoting fitting errors are covered with markers. In this numerical simulation, the angle~$\theta$ and widths~$(\sigma_{long},\sigma_{short})$ of the initial distribution of atoms are fixed at $\theta = 0^{\circ}$ and $(\sigma_{long},\sigma_{short}) =(7.50,3.75)~\mu{\rm m}$, respectively.}
\label{figureS8}
\end{center}
\end{figure}

To support our spin domains scenario, we numerically simulated the TOF of samples with spin domains. To simplify the numerical simulation, we assumed that Gaussian-shaped wavefunctions were located at each lattice site. The width of the wavefunction was extracted from the Gaussian fit to the Wannier function in the $ab \; initio$ calculation with a lattice depth $V_0=3.0~E_R$. The phases of Sp1 or Sp2 configurations for a homogeneous system were attached to the wavefunctions. We set domain walls perpendicular to the long axis of the elongated initial atom distribution and the opposite chiral modes for each region (Fig.~\ref{figureS8}(a)). The weight of the wavefunctions was determined by the initial density profile approximated by the 2D elliptic Gaussian function with the tilt angle (Fig.~\ref{figureS7}(a)). We obtained the density profiles after the TOF of 5 ms by summing all wavefunctions after free expansion. Here, we neglect the effect of residual trapping confinement owing to the vertical lattice. 

The simulated density distributions with the number of domain walls from 1 to 3 are presented in Fig.~\ref{figureS8}(b). In the case of a single domain wall, one can see the effect of the interference between the two chiral modes. When multiple domain walls are formed, on the other hand, interference between the same chiral modes also occurs. We also checked the fast Fourier transformation of the density distributions, as we conducted for the experimental data in the main text, and observed that high-frequency components appeared along the long axis of the initial atom distributions (Fig.~\ref{figureS8}(c)). To quantify the FFT signals, we made a fit to the data in the same way as is described in the previous section~\ref{sec6}. Since the numerical simulation is free from noise, we omitted Gauss3 in Eq.~\ref{eq:TriGauss}. Figure~\ref{figureS8}(d) shows the extracted widths in FFT space with various numbers of domain walls. The results tell us that the larger the number of domain walls, the larger the aspect of the FFT signals.

\begin{thebibliography}{40}%
\makeatletter
\providecommand \@ifxundefined [1]{%
 \@ifx{#1\undefined}
}%
\providecommand \@ifnum [1]{%
 \ifnum #1\expandafter \@firstoftwo
 \else \expandafter \@secondoftwo
 \fi
}%
\providecommand \@ifx [1]{%
 \ifx #1\expandafter \@firstoftwo
 \else \expandafter \@secondoftwo
 \fi
}%
\providecommand \natexlab [1]{#1}%
\providecommand \enquote  [1]{``#1''}%
\providecommand \bibnamefont  [1]{#1}%
\providecommand \bibfnamefont [1]{#1}%
\providecommand \citenamefont [1]{#1}%
\providecommand \href@noop [0]{\@secondoftwo}%
\providecommand \href [0]{\begingroup \@sanitize@url \@href}%
\providecommand \@href[1]{\@@startlink{#1}\@@href}%
\providecommand \@@href[1]{\endgroup#1\@@endlink}%
\providecommand \@sanitize@url [0]{\catcode `\\12\catcode `\$12\catcode
  `\&12\catcode `\#12\catcode `\^12\catcode `\_12\catcode `\%12\relax}%
\providecommand \@@startlink[1]{}%
\providecommand \@@endlink[0]{}%
\providecommand \url  [0]{\begingroup\@sanitize@url \@url }%
\providecommand \@url [1]{\endgroup\@href {#1}{\urlprefix }}%
\providecommand \urlprefix  [0]{URL }%
\providecommand \Eprint [0]{\href }%
\providecommand \doibase [0]{http://dx.doi.org/}%
\providecommand \selectlanguage [0]{\@gobble}%
\providecommand \bibinfo  [0]{\@secondoftwo}%
\providecommand \bibfield  [0]{\@secondoftwo}%
\providecommand \translation [1]{[#1]}%
\providecommand \BibitemOpen [0]{}%
\providecommand \bibitemStop [0]{}%
\providecommand \bibitemNoStop [0]{.\EOS\space}%
\providecommand \EOS [0]{\spacefactor3000\relax}%
\providecommand \BibitemShut  [1]{\csname bibitem#1\endcsname}%
\let\auto@bib@innerbib\@empty
\bibitem [{\citenamefont {{H. Diep}}(2004)}]{Diep:2004}%
  \BibitemOpen
  \bibfield  {author} {\bibinfo {author} {\bibnamefont {{H. Diep}}},\
  }\href@noop {} {\emph {\bibinfo {title} {{Frustrated Spin Systems}}}}\
  (\bibinfo  {publisher} {World Scientific, Singapore},\ \bibinfo {year}
  {2004})\BibitemShut {NoStop}%
\bibitem [{\citenamefont {Moessner}\ and\ \citenamefont
  {Ramirez}(2006)}]{Moessner:2006}%
  \BibitemOpen
  \bibfield  {author} {\bibinfo {author} {\bibfnamefont {R.}~\bibnamefont
  {Moessner}}\ and\ \bibinfo {author} {\bibfnamefont {A.~P.}\ \bibnamefont
  {Ramirez}},\ }\href {https://doi.org/10.1063/1.2186278} {\bibfield  {journal}
  {\bibinfo  {journal} {Physics Today}\ }\textbf {\bibinfo {volume} {59}},\
  \bibinfo {pages} {24} (\bibinfo {year} {2006})}\BibitemShut {NoStop}%
\bibitem [{\citenamefont {Balents}(2010)}]{Balents:2010}%
  \BibitemOpen
  \bibfield  {author} {\bibinfo {author} {\bibfnamefont {L.}~\bibnamefont
  {Balents}},\ }\href {\doibase 10.1038/nature08917} {\bibfield  {journal}
  {\bibinfo  {journal} {Nature}\ }\textbf {\bibinfo {volume} {464}},\ \bibinfo
  {pages} {199} (\bibinfo {year} {2010})}\BibitemShut {NoStop}%
\bibitem [{\citenamefont {Harrison}(2004)}]{HarrisonA2004}%
  \BibitemOpen
  \bibfield  {author} {\bibinfo {author} {\bibfnamefont {A.}~\bibnamefont
  {Harrison}},\ }\href {\doibase 10.1088/0953-8984/16/11/001} {\bibfield
  {journal} {\bibinfo  {journal} {Journal of Physics: Condensed Matter}\
  }\textbf {\bibinfo {volume} {16}},\ \bibinfo {pages} {S553} (\bibinfo {year}
  {2004})}\BibitemShut {NoStop}%
\bibitem [{\citenamefont {Kim}\ \emph {et~al.}(2010)\citenamefont {Kim},
  \citenamefont {Chang}, \citenamefont {Korenblit}, \citenamefont {Islam},
  \citenamefont {Edwards}, \citenamefont {Freericks}, \citenamefont {Lin},
  \citenamefont {Duan},\ and\ \citenamefont {Monroe}}]{Kim:2010}%
  \BibitemOpen
  \bibfield  {author} {\bibinfo {author} {\bibfnamefont {K.}~\bibnamefont
  {Kim}}, \bibinfo {author} {\bibfnamefont {M.~S.}\ \bibnamefont {Chang}},
  \bibinfo {author} {\bibfnamefont {S.}~\bibnamefont {Korenblit}}, \bibinfo
  {author} {\bibfnamefont {R.}~\bibnamefont {Islam}}, \bibinfo {author}
  {\bibfnamefont {E.~E.}\ \bibnamefont {Edwards}}, \bibinfo {author}
  {\bibfnamefont {J.~K.}\ \bibnamefont {Freericks}}, \bibinfo {author}
  {\bibfnamefont {G.~D.}\ \bibnamefont {Lin}}, \bibinfo {author} {\bibfnamefont
  {L.~M.}\ \bibnamefont {Duan}}, \ and\ \bibinfo {author} {\bibfnamefont
  {C.}~\bibnamefont {Monroe}},\ }\href {\doibase 10.1038/nature09071}
  {\bibfield  {journal} {\bibinfo  {journal} {Nature}\ }\textbf {\bibinfo
  {volume} {465}},\ \bibinfo {pages} {590} (\bibinfo {year}
  {2010})}\BibitemShut {NoStop}%
\bibitem [{\citenamefont {Qiao}\ \emph {et~al.}(2022)\citenamefont {Qiao},
  \citenamefont {Cai}, \citenamefont {Wang}, \citenamefont {Du}, \citenamefont
  {Jin}, \citenamefont {Chen}, \citenamefont {Wang}, \citenamefont {Luan},
  \citenamefont {Gao}, \citenamefont {Sun}, \citenamefont {Tian}, \citenamefont
  {Zhang},\ and\ \citenamefont {Kim}}]{Qiao:2022}%
  \BibitemOpen
  \bibfield  {author} {\bibinfo {author} {\bibfnamefont {M.}~\bibnamefont
  {Qiao}}, \bibinfo {author} {\bibfnamefont {Z.}~\bibnamefont {Cai}}, \bibinfo
  {author} {\bibfnamefont {Y.}~\bibnamefont {Wang}}, \bibinfo {author}
  {\bibfnamefont {B.}~\bibnamefont {Du}}, \bibinfo {author} {\bibfnamefont
  {N.}~\bibnamefont {Jin}}, \bibinfo {author} {\bibfnamefont {W.}~\bibnamefont
  {Chen}}, \bibinfo {author} {\bibfnamefont {P.}~\bibnamefont {Wang}}, \bibinfo
  {author} {\bibfnamefont {C.}~\bibnamefont {Luan}}, \bibinfo {author}
  {\bibfnamefont {E.}~\bibnamefont {Gao}}, \bibinfo {author} {\bibfnamefont
  {X.}~\bibnamefont {Sun}}, \bibinfo {author} {\bibfnamefont {H.}~\bibnamefont
  {Tian}}, \bibinfo {author} {\bibfnamefont {J.}~\bibnamefont {Zhang}}, \ and\
  \bibinfo {author} {\bibfnamefont {K.}~\bibnamefont {Kim}},\ }\href
  {http://arxiv.org/abs/2204.07283} {\bibfield  {journal} {\bibinfo  {journal}
  {arXiv:2204.07283}\ } (\bibinfo {year} {2022})}\BibitemShut {NoStop}%
\bibitem [{\citenamefont {Struck}\ \emph {et~al.}(2011)\citenamefont {Struck},
  \citenamefont {{\"{O}}lschl{\"{a}}ger}, \citenamefont {Le~Targat},
  \citenamefont {Soltan-Panahi}, \citenamefont {Eckardt}, \citenamefont
  {Lewenstein}, \citenamefont {Windpassinger},\ and\ \citenamefont
  {Sengstock}}]{Struck:2011}%
  \BibitemOpen
  \bibfield  {author} {\bibinfo {author} {\bibfnamefont {J.}~\bibnamefont
  {Struck}}, \bibinfo {author} {\bibfnamefont {C.}~\bibnamefont
  {{\"{O}}lschl{\"{a}}ger}}, \bibinfo {author} {\bibfnamefont {R.}~\bibnamefont
  {Le~Targat}}, \bibinfo {author} {\bibfnamefont {P.}~\bibnamefont
  {Soltan-Panahi}}, \bibinfo {author} {\bibfnamefont {A.}~\bibnamefont
  {Eckardt}}, \bibinfo {author} {\bibfnamefont {M.}~\bibnamefont {Lewenstein}},
  \bibinfo {author} {\bibfnamefont {P.}~\bibnamefont {Windpassinger}}, \ and\
  \bibinfo {author} {\bibfnamefont {K.}~\bibnamefont {Sengstock}},\ }\href
  {\doibase 10.1126/science.1207239} {\bibfield  {journal} {\bibinfo  {journal}
  {Science}\ }\textbf {\bibinfo {volume} {333}},\ \bibinfo {pages} {996}
  (\bibinfo {year} {2011})}\BibitemShut {NoStop}%
\bibitem [{\citenamefont {Mongkolkiattichai}\ \emph {et~al.}(2022)\citenamefont
  {Mongkolkiattichai}, \citenamefont {Liu}, \citenamefont {Garwood},
  \citenamefont {Yang},\ and\ \citenamefont
  {Schauss}}]{Mongkolkiattichai:2022}%
  \BibitemOpen
  \bibfield  {author} {\bibinfo {author} {\bibfnamefont {J.}~\bibnamefont
  {Mongkolkiattichai}}, \bibinfo {author} {\bibfnamefont {L.}~\bibnamefont
  {Liu}}, \bibinfo {author} {\bibfnamefont {D.}~\bibnamefont {Garwood}},
  \bibinfo {author} {\bibfnamefont {J.}~\bibnamefont {Yang}}, \ and\ \bibinfo
  {author} {\bibfnamefont {P.}~\bibnamefont {Schauss}},\ }\href
  {http://arxiv.org/abs/2210.14895} {\bibfield  {journal} {\bibinfo  {journal}
  {arXiv:2210.14895}\ } (\bibinfo {year} {2022})}\BibitemShut {NoStop}%
\bibitem [{\citenamefont {Xu}\ \emph {et~al.}(2023)\citenamefont {Xu},
  \citenamefont {Kendrick}, \citenamefont {Kale}, \citenamefont {Gang},
  \citenamefont {Ji}, \citenamefont {Scalettar}, \citenamefont {Lebrat},\ and\
  \citenamefont {Greiner}}]{XuM2023}%
  \BibitemOpen
  \bibfield  {author} {\bibinfo {author} {\bibfnamefont {M.}~\bibnamefont
  {Xu}}, \bibinfo {author} {\bibfnamefont {L.~H.}\ \bibnamefont {Kendrick}},
  \bibinfo {author} {\bibfnamefont {A.}~\bibnamefont {Kale}}, \bibinfo {author}
  {\bibfnamefont {Y.}~\bibnamefont {Gang}}, \bibinfo {author} {\bibfnamefont
  {G.}~\bibnamefont {Ji}}, \bibinfo {author} {\bibfnamefont {R.~T.}\
  \bibnamefont {Scalettar}}, \bibinfo {author} {\bibfnamefont {M.}~\bibnamefont
  {Lebrat}}, \ and\ \bibinfo {author} {\bibfnamefont {M.}~\bibnamefont
  {Greiner}},\ }\href {\doibase 10.1038/s41586-023-06280-5} {\bibfield
  {journal} {\bibinfo  {journal} {Nature}\ } (\bibinfo {year} {2023}),\
  10.1038/s41586-023-06280-5}\BibitemShut {NoStop}%
\bibitem [{\citenamefont {Lebrat}\ \emph {et~al.}(2023)\citenamefont {Lebrat},
  \citenamefont {Xu}, \citenamefont {Kendrick}, \citenamefont {Kale},
  \citenamefont {Gang}, \citenamefont {Seetharaman}, \citenamefont {Morera},
  \citenamefont {Khatami}, \citenamefont {Demler},\ and\ \citenamefont
  {Greiner}}]{LebratM2023}%
  \BibitemOpen
  \bibfield  {author} {\bibinfo {author} {\bibfnamefont {M.}~\bibnamefont
  {Lebrat}}, \bibinfo {author} {\bibfnamefont {M.}~\bibnamefont {Xu}}, \bibinfo
  {author} {\bibfnamefont {L.~H.}\ \bibnamefont {Kendrick}}, \bibinfo {author}
  {\bibfnamefont {A.}~\bibnamefont {Kale}}, \bibinfo {author} {\bibfnamefont
  {Y.}~\bibnamefont {Gang}}, \bibinfo {author} {\bibfnamefont {P.}~\bibnamefont
  {Seetharaman}}, \bibinfo {author} {\bibfnamefont {I.}~\bibnamefont {Morera}},
  \bibinfo {author} {\bibfnamefont {E.}~\bibnamefont {Khatami}}, \bibinfo
  {author} {\bibfnamefont {E.}~\bibnamefont {Demler}}, \ and\ \bibinfo {author}
  {\bibfnamefont {M.}~\bibnamefont {Greiner}},\ }\href
  {http://arxiv.org/abs/2308.12269} {\bibinfo {journal}{arXiv:2308.12269} (\bibinfo {year} {2023})}\BibitemShut
  {NoStop}%
\bibitem [{\citenamefont {Prichard}\ \emph {et~al.}(2023)\citenamefont
  {Prichard}, \citenamefont {Spar}, \citenamefont {Morera}, \citenamefont
  {Demler}, \citenamefont {Yan},\ and\ \citenamefont {Bakr}}]{PrichardM2023}%
  \BibitemOpen
  \bibfield  {author} {\bibinfo {author} {\bibfnamefont {M.~L.}\ \bibnamefont
  {Prichard}}, \bibinfo {author} {\bibfnamefont {B.~M.}\ \bibnamefont {Spar}},
  \bibinfo {author} {\bibfnamefont {I.}~\bibnamefont {Morera}}, \bibinfo
  {author} {\bibfnamefont {E.}~\bibnamefont {Demler}}, \bibinfo {author}
  {\bibfnamefont {Z.~Z.}\ \bibnamefont {Yan}}, \ and\ \bibinfo {author}
  {\bibfnamefont {W.~S.}\ \bibnamefont {Bakr}},\ }\href
  {http://arxiv.org/abs/2308.12951} {\bibfield  {journal} {\bibinfo  {journal}
  {arXiv:2308.12951}\ } (\bibinfo {year} {2023})}\BibitemShut {NoStop}%
\bibitem [{\citenamefont {Scholl}\ \emph {et~al.}(2021)\citenamefont {Scholl},
  \citenamefont {Schuler}, \citenamefont {Williams}, \citenamefont
  {Eberharter}, \citenamefont {Barredo}, \citenamefont {Schymik}, \citenamefont
  {Lienhard}, \citenamefont {Henry}, \citenamefont {Lang}, \citenamefont
  {Lahaye}, \citenamefont {L{\"{a}}uchli},\ and\ \citenamefont
  {Browaeys}}]{Scholl:2021}%
  \BibitemOpen
  \bibfield  {author} {\bibinfo {author} {\bibfnamefont {P.}~\bibnamefont
  {Scholl}}, \bibinfo {author} {\bibfnamefont {M.}~\bibnamefont {Schuler}},
  \bibinfo {author} {\bibfnamefont {H.~J.}\ \bibnamefont {Williams}}, \bibinfo
  {author} {\bibfnamefont {A.~A.}\ \bibnamefont {Eberharter}}, \bibinfo
  {author} {\bibfnamefont {D.}~\bibnamefont {Barredo}}, \bibinfo {author}
  {\bibfnamefont {K.~N.}\ \bibnamefont {Schymik}}, \bibinfo {author}
  {\bibfnamefont {V.}~\bibnamefont {Lienhard}}, \bibinfo {author}
  {\bibfnamefont {L.~P.}\ \bibnamefont {Henry}}, \bibinfo {author}
  {\bibfnamefont {T.~C.}\ \bibnamefont {Lang}}, \bibinfo {author}
  {\bibfnamefont {T.}~\bibnamefont {Lahaye}}, \bibinfo {author} {\bibfnamefont
  {A.~M.}\ \bibnamefont {L{\"{a}}uchli}}, \ and\ \bibinfo {author}
  {\bibfnamefont {A.}~\bibnamefont {Browaeys}},\ }\href {\doibase
  10.1038/s41586-021-03585-1} {\bibfield  {journal} {\bibinfo  {journal}
  {Nature}\ }\textbf {\bibinfo {volume} {595}},\ \bibinfo {pages} {233}
  (\bibinfo {year} {2021})}\BibitemShut {NoStop}%
\bibitem [{\citenamefont {Semeghini}\ \emph {et~al.}(2021)\citenamefont
  {Semeghini}, \citenamefont {Levine}, \citenamefont {Keesling}, \citenamefont
  {Ebadi}, \citenamefont {Wang}, \citenamefont {Bluvstein}, \citenamefont
  {Verresen}, \citenamefont {Pichler}, \citenamefont {Kalinowski},
  \citenamefont {Samajdar}, \citenamefont {Omran}, \citenamefont {Sachdev},
  \citenamefont {Vishwanath}, \citenamefont {Greiner}, \citenamefont
  {Vuleti{\'{c}}},\ and\ \citenamefont {Lukin}}]{Semeghini:2021}%
  \BibitemOpen
  \bibfield  {author} {\bibinfo {author} {\bibfnamefont {G.}~\bibnamefont
  {Semeghini}}, \bibinfo {author} {\bibfnamefont {H.}~\bibnamefont {Levine}},
  \bibinfo {author} {\bibfnamefont {A.}~\bibnamefont {Keesling}}, \bibinfo
  {author} {\bibfnamefont {S.}~\bibnamefont {Ebadi}}, \bibinfo {author}
  {\bibfnamefont {T.~T.}\ \bibnamefont {Wang}}, \bibinfo {author}
  {\bibfnamefont {D.}~\bibnamefont {Bluvstein}}, \bibinfo {author}
  {\bibfnamefont {R.}~\bibnamefont {Verresen}}, \bibinfo {author}
  {\bibfnamefont {H.}~\bibnamefont {Pichler}}, \bibinfo {author} {\bibfnamefont
  {M.}~\bibnamefont {Kalinowski}}, \bibinfo {author} {\bibfnamefont
  {R.}~\bibnamefont {Samajdar}}, \bibinfo {author} {\bibfnamefont
  {A.}~\bibnamefont {Omran}}, \bibinfo {author} {\bibfnamefont
  {S.}~\bibnamefont {Sachdev}}, \bibinfo {author} {\bibfnamefont
  {A.}~\bibnamefont {Vishwanath}}, \bibinfo {author} {\bibfnamefont
  {M.}~\bibnamefont {Greiner}}, \bibinfo {author} {\bibfnamefont
  {V.}~\bibnamefont {Vuleti{\'{c}}}}, \ and\ \bibinfo {author} {\bibfnamefont
  {M.~D.}\ \bibnamefont {Lukin}},\ }\href {\doibase 10.1126/science.abi8794}
  {\bibfield  {journal} {\bibinfo  {journal} {Science}\ }\textbf {\bibinfo
  {volume} {374}},\ \bibinfo {pages} {1242} (\bibinfo {year}
  {2021})}\BibitemShut {NoStop}%
\bibitem [{\citenamefont {King}\ \emph {et~al.}(2021)\citenamefont {King},
  \citenamefont {Nisoli}, \citenamefont {Dahl}, \citenamefont
  {Poulin-Lamarre},\ and\ \citenamefont {Lopez-Bezanilla}}]{King:2021}%
  \BibitemOpen
  \bibfield  {author} {\bibinfo {author} {\bibfnamefont {A.~D.}\ \bibnamefont
  {King}}, \bibinfo {author} {\bibfnamefont {C.}~\bibnamefont {Nisoli}},
  \bibinfo {author} {\bibfnamefont {E.~D.}\ \bibnamefont {Dahl}}, \bibinfo
  {author} {\bibfnamefont {G.}~\bibnamefont {Poulin-Lamarre}}, \ and\ \bibinfo
  {author} {\bibfnamefont {A.}~\bibnamefont {Lopez-Bezanilla}},\ }\href
  {\doibase 10.1126/science.abe2824} {\bibfield  {journal} {\bibinfo  {journal}
  {Science}\ }\textbf {\bibinfo {volume} {373}},\ \bibinfo {pages} {576}
  (\bibinfo {year} {2021})}\BibitemShut {NoStop}%
\bibitem [{\citenamefont {Cosmic}\ \emph {et~al.}(2020)\citenamefont {Cosmic},
  \citenamefont {Kawabata}, \citenamefont {Ashida}, \citenamefont {Ikegami},
  \citenamefont {Furukawa}, \citenamefont {Patil}, \citenamefont {Taylor},\
  and\ \citenamefont {Nakamura}}]{Cosmic:2020}%
  \BibitemOpen
  \bibfield  {author} {\bibinfo {author} {\bibfnamefont {R.}~\bibnamefont
  {Cosmic}}, \bibinfo {author} {\bibfnamefont {K.}~\bibnamefont {Kawabata}},
  \bibinfo {author} {\bibfnamefont {Y.}~\bibnamefont {Ashida}}, \bibinfo
  {author} {\bibfnamefont {H.}~\bibnamefont {Ikegami}}, \bibinfo {author}
  {\bibfnamefont {S.}~\bibnamefont {Furukawa}}, \bibinfo {author}
  {\bibfnamefont {P.}~\bibnamefont {Patil}}, \bibinfo {author} {\bibfnamefont
  {J.~M.}\ \bibnamefont {Taylor}}, \ and\ \bibinfo {author} {\bibfnamefont
  {Y.}~\bibnamefont {Nakamura}},\ }\href {\doibase 10.1103/PhysRevB.102.094509}
  {\bibfield  {journal} {\bibinfo  {journal} {Physical Review B}\ }\textbf
  {\bibinfo {volume} {102}},\ \bibinfo {pages} {094509} (\bibinfo {year}
  {2020})}\BibitemShut {NoStop}%
\bibitem [{\citenamefont {Teitel}\ and\ \citenamefont
  {Jayaprakash}(1983)}]{Teitel:1983}%
  \BibitemOpen
  \bibfield  {author} {\bibinfo {author} {\bibfnamefont {S.}~\bibnamefont
  {Teitel}}\ and\ \bibinfo {author} {\bibfnamefont {C.}~\bibnamefont
  {Jayaprakash}},\ }\href {\doibase 10.1103/PhysRevB.27.598} {\bibfield
  {journal} {\bibinfo  {journal} {Physical Review B}\ }\textbf {\bibinfo
  {volume} {27}},\ \bibinfo {pages} {598} (\bibinfo {year} {1983})}\BibitemShut
  {NoStop}%
\bibitem [{\citenamefont {Miyashita}\ and\ \citenamefont
  {Shiba}(1984)}]{Miyashita:1984}%
  \BibitemOpen
  \bibfield  {author} {\bibinfo {author} {\bibfnamefont {S.}~\bibnamefont
  {Miyashita}}\ and\ \bibinfo {author} {\bibfnamefont {H.}~\bibnamefont
  {Shiba}},\ }\href@noop {} {\bibfield  {journal} {\bibinfo  {journal} {Journal
  of the Physical Society of Japan}\ }\textbf {\bibinfo {volume} {53}},\
  \bibinfo {pages} {1145} (\bibinfo {year} {1984})}\BibitemShut {NoStop}%
\bibitem [{\citenamefont {Lee}\ \emph {et~al.}(1984)\citenamefont {Lee},
  \citenamefont {Joannopoulos}, \citenamefont {Negele},\ and\ \citenamefont
  {Landau}}]{Lee:1984}%
  \BibitemOpen
  \bibfield  {author} {\bibinfo {author} {\bibfnamefont {D.~H.}\ \bibnamefont
  {Lee}}, \bibinfo {author} {\bibfnamefont {J.~D.}\ \bibnamefont
  {Joannopoulos}}, \bibinfo {author} {\bibfnamefont {J.~W.}\ \bibnamefont
  {Negele}}, \ and\ \bibinfo {author} {\bibfnamefont {D.~P.}\ \bibnamefont
  {Landau}},\ }\href {\doibase 10.1103/PhysRevLett.52.433} {\bibfield
  {journal} {\bibinfo  {journal} {Physical Review Letters}\ }\textbf {\bibinfo
  {volume} {52}},\ \bibinfo {pages} {433} (\bibinfo {year} {1984})}\BibitemShut
  {NoStop}%
\bibitem [{\citenamefont {Song}\ and\ \citenamefont {Zhang}(2022)}]{Song:2022}%
  \BibitemOpen
  \bibfield  {author} {\bibinfo {author} {\bibfnamefont {F.~F.}\ \bibnamefont
  {Song}}\ and\ \bibinfo {author} {\bibfnamefont {G.~M.}\ \bibnamefont
  {Zhang}},\ }\href {\doibase 10.1103/PhysRevB.105.134516} {\bibfield
  {journal} {\bibinfo  {journal} {Physical Review B}\ }\textbf {\bibinfo
  {volume} {105}},\ \bibinfo {pages} {134516} (\bibinfo {year}
  {2022})}\BibitemShut {NoStop}%
\bibitem [{\citenamefont {Obuchi}\ and\ \citenamefont
  {Kawamura}(2012)}]{Obuchi:2012}%
  \BibitemOpen
  \bibfield  {author} {\bibinfo {author} {\bibfnamefont {T.}~\bibnamefont
  {Obuchi}}\ and\ \bibinfo {author} {\bibfnamefont {H.}~\bibnamefont
  {Kawamura}},\ }\href {\doibase 10.1143/JPSJ.81.054003} {\bibfield  {journal}
  {\bibinfo  {journal} {Journal of the Physical Society of Japan}\ }\textbf
  {\bibinfo {volume} {81}},\ \bibinfo {pages} {054003} (\bibinfo {year}
  {2012})}\BibitemShut {NoStop}%
\bibitem [{\citenamefont {Lv}\ \emph {et~al.}(2013)\citenamefont {Lv},
  \citenamefont {Garoni},\ and\ \citenamefont {Deng}}]{Lv:2013}%
  \BibitemOpen
  \bibfield  {author} {\bibinfo {author} {\bibfnamefont {J.~P.}\ \bibnamefont
  {Lv}}, \bibinfo {author} {\bibfnamefont {T.~M.}\ \bibnamefont {Garoni}}, \
  and\ \bibinfo {author} {\bibfnamefont {Y.}~\bibnamefont {Deng}},\ }\href
  {\doibase 10.1103/PhysRevB.87.024108} {\bibfield  {journal} {\bibinfo
  {journal} {Physical Review B}\ }\textbf {\bibinfo {volume} {87}},\ \bibinfo
  {pages} {024108} (\bibinfo {year} {2013})}\BibitemShut {NoStop}%
\bibitem [{\citenamefont {Eckardt}(2017)}]{RevModPhys.89.011004}%
  \BibitemOpen
  \bibfield  {author} {\bibinfo {author} {\bibfnamefont {A.}~\bibnamefont
  {Eckardt}},\ }\href {\doibase 10.1103/RevModPhys.89.011004} {\bibfield
  {journal} {\bibinfo  {journal} {Reviews of Modern Physics}\ }\textbf
  {\bibinfo {volume} {89}},\ \bibinfo {pages} {011004} (\bibinfo {year}
  {2017})}\BibitemShut {NoStop}%
\bibitem [{\citenamefont {Wang}\ \emph {et~al.}(2009)\citenamefont {Wang},
  \citenamefont {Pollmann},\ and\ \citenamefont {Vishwanath}}]{Wang:2009}%
  \BibitemOpen
  \bibfield  {author} {\bibinfo {author} {\bibfnamefont {F.}~\bibnamefont
  {Wang}}, \bibinfo {author} {\bibfnamefont {F.}~\bibnamefont {Pollmann}}, \
  and\ \bibinfo {author} {\bibfnamefont {A.}~\bibnamefont {Vishwanath}},\
  }\href {\doibase 10.1103/PhysRevLett.102.017203} {\bibfield  {journal}
  {\bibinfo  {journal} {Physical Review Letters}\ }\textbf {\bibinfo {volume}
  {102}},\ \bibinfo {pages} {017203} (\bibinfo {year} {2009})}\BibitemShut
  {NoStop}%
\bibitem [{\citenamefont {Jiang}\ \emph {et~al.}(2009)\citenamefont {Jiang},
  \citenamefont {Weng}, \citenamefont {Weng}, \citenamefont {Sheng},\ and\
  \citenamefont {Balents}}]{Jian:2009}%
  \BibitemOpen
  \bibfield  {author} {\bibinfo {author} {\bibfnamefont {H.~C.}\ \bibnamefont
  {Jiang}}, \bibinfo {author} {\bibfnamefont {M.~Q.}\ \bibnamefont {Weng}},
  \bibinfo {author} {\bibfnamefont {Z.~Y.}\ \bibnamefont {Weng}}, \bibinfo
  {author} {\bibfnamefont {D.~N.}\ \bibnamefont {Sheng}}, \ and\ \bibinfo
  {author} {\bibfnamefont {L.}~\bibnamefont {Balents}},\ }\href {\doibase
  10.1103/PhysRevB.79.020409} {\bibfield  {journal} {\bibinfo  {journal}
  {Physical Review B}\ }\textbf {\bibinfo {volume} {79}},\ \bibinfo {pages}
  {020409(R)} (\bibinfo {year} {2009})}\BibitemShut {NoStop}%
\bibitem [{\citenamefont {Heidarian}\ and\ \citenamefont
  {Paramekanti}(2010)}]{Heidarian:2010}%
  \BibitemOpen
  \bibfield  {author} {\bibinfo {author} {\bibfnamefont {D.}~\bibnamefont
  {Heidarian}}\ and\ \bibinfo {author} {\bibfnamefont {A.}~\bibnamefont
  {Paramekanti}},\ }\href {\doibase 10.1103/PhysRevLett.104.015301} {\bibfield
  {journal} {\bibinfo  {journal} {Physical Review Letters}\ }\textbf {\bibinfo
  {volume} {104}},\ \bibinfo {pages} {015301} (\bibinfo {year}
  {2010})}\BibitemShut {NoStop}%
\bibitem [{sup()}]{supp_Lieb}%
  \BibitemOpen
  \href@noop {} {\bibinfo  {journal} {See Supplemental Material at [url] for
  the theoretical and experimental details, which includes Refs.~\cite{RevModPhys.89.011004, Struck:2011, Goldman:2014, Guo:2019, Arlinghaus:2010, Weinberg:2015, Martikainen:2011, Paul:2013, PhysRevLett.119.200402}}\
  }\BibitemShut {NoStop}%
\bibitem [{\citenamefont {Weinberg}\ \emph {et~al.}(2015)\citenamefont
  {Weinberg}, \citenamefont {{\"{O}}lschl{\"{a}}ger}, \citenamefont
  {Str{\"{a}}ter}, \citenamefont {Prelle}, \citenamefont {Eckardt},
  \citenamefont {Sengstock},\ and\ \citenamefont {Simonet}}]{Weinberg:2015}%
  \BibitemOpen
\bibfield  {journal} {  }\bibfield  {author} {\bibinfo {author} {\bibfnamefont
  {M.}~\bibnamefont {Weinberg}}, \bibinfo {author} {\bibfnamefont
  {C.}~\bibnamefont {{\"{O}}lschl{\"{a}}ger}}, \bibinfo {author} {\bibfnamefont
  {C.}~\bibnamefont {Str{\"{a}}ter}}, \bibinfo {author} {\bibfnamefont
  {S.}~\bibnamefont {Prelle}}, \bibinfo {author} {\bibfnamefont
  {A.}~\bibnamefont {Eckardt}}, \bibinfo {author} {\bibfnamefont
  {K.}~\bibnamefont {Sengstock}}, \ and\ \bibinfo {author} {\bibfnamefont
  {J.}~\bibnamefont {Simonet}},\ }\href {\doibase 10.1103/PhysRevA.92.043621}
  {\bibfield  {journal} {\bibinfo  {journal} {Phys. Rev. A}\ }\textbf {\bibinfo
  {volume} {92}},\ \bibinfo {pages} {043621} (\bibinfo {year}
  {2015})}\BibitemShut {NoStop}%
\bibitem [{\citenamefont {Reitter}\ \emph {et~al.}(2017)\citenamefont
  {Reitter}, \citenamefont {N{\"{a}}ger}, \citenamefont {Wintersperger},
  \citenamefont {Str{\"{a}}ter}, \citenamefont {Bloch}, \citenamefont
  {Eckardt},\ and\ \citenamefont {Schneider}}]{PhysRevLett.119.200402}%
  \BibitemOpen
  \bibfield  {author} {\bibinfo {author} {\bibfnamefont {M.}~\bibnamefont
  {Reitter}}, \bibinfo {author} {\bibfnamefont {J.}~\bibnamefont
  {N{\"{a}}ger}}, \bibinfo {author} {\bibfnamefont {K.}~\bibnamefont
  {Wintersperger}}, \bibinfo {author} {\bibfnamefont {C.}~\bibnamefont
  {Str{\"{a}}ter}}, \bibinfo {author} {\bibfnamefont {I.}~\bibnamefont
  {Bloch}}, \bibinfo {author} {\bibfnamefont {A.}~\bibnamefont {Eckardt}}, \
  and\ \bibinfo {author} {\bibfnamefont {U.}~\bibnamefont {Schneider}},\ }\href
  {\doibase 10.1103/PhysRevLett.119.200402} {\bibfield  {journal} {\bibinfo
  {journal} {Physical Review Letters}\ }\textbf {\bibinfo {volume} {119}},\
  \bibinfo {pages} {200402} (\bibinfo {year} {2017})}\BibitemShut {NoStop}%
\bibitem [{\citenamefont {Bakr}\ \emph {et~al.}(2010)\citenamefont {Bakr},
  \citenamefont {Peng}, \citenamefont {Tai}, \citenamefont {Ma}, \citenamefont
  {Simon}, \citenamefont {Gillen}, \citenamefont {F{\"{o}}lling}, \citenamefont
  {Pollet},\ and\ \citenamefont {Greiner}}]{Bakr:2010}%
  \BibitemOpen
  \bibfield  {author} {\bibinfo {author} {\bibfnamefont {W.~S.}\ \bibnamefont
  {Bakr}}, \bibinfo {author} {\bibfnamefont {A.}~\bibnamefont {Peng}}, \bibinfo
  {author} {\bibfnamefont {M.~E.}\ \bibnamefont {Tai}}, \bibinfo {author}
  {\bibfnamefont {R.}~\bibnamefont {Ma}}, \bibinfo {author} {\bibfnamefont
  {J.}~\bibnamefont {Simon}}, \bibinfo {author} {\bibfnamefont {J.~I.}\
  \bibnamefont {Gillen}}, \bibinfo {author} {\bibfnamefont {S.}~\bibnamefont
  {F{\"{o}}lling}}, \bibinfo {author} {\bibfnamefont {L.}~\bibnamefont
  {Pollet}}, \ and\ \bibinfo {author} {\bibfnamefont {M.}~\bibnamefont
  {Greiner}},\ }\href
  {http://science.sciencemag.org/content/329/5991/547.abstract} {\bibfield
  {journal} {\bibinfo  {journal} {Science}\ }\textbf {\bibinfo {volume}
  {329}},\ \bibinfo {pages} {547} (\bibinfo {year} {2010})}\BibitemShut
  {NoStop}%
\bibitem [{\citenamefont {Yamamoto}\ \emph
  {et~al.}(2020{\natexlab{a}})\citenamefont {Yamamoto}, \citenamefont {Ozawa},
  \citenamefont {Nak}, \citenamefont {Nakamura},\ and\ \citenamefont
  {Fukuhara}}]{RYamamoto:2020}%
  \BibitemOpen
  \bibfield  {author} {\bibinfo {author} {\bibfnamefont {R.}~\bibnamefont
  {Yamamoto}}, \bibinfo {author} {\bibfnamefont {H.}~\bibnamefont {Ozawa}},
  \bibinfo {author} {\bibfnamefont {D.~C.}\ \bibnamefont {Nak}}, \bibinfo
  {author} {\bibfnamefont {I.}~\bibnamefont {Nakamura}}, \ and\ \bibinfo
  {author} {\bibfnamefont {T.}~\bibnamefont {Fukuhara}},\ }\href {\doibase
  10.1088/1367-2630/abcdc8} {\bibfield  {journal} {\bibinfo  {journal} {New
  Journal of Physics}\ }\textbf {\bibinfo {volume} {22}},\ \bibinfo {pages}
  {123028} (\bibinfo {year} {2020}{\natexlab{a}})}\BibitemShut {NoStop}%
\bibitem [{\citenamefont {Eckardt}\ \emph {et~al.}(2010)\citenamefont
  {Eckardt}, \citenamefont {Hauke}, \citenamefont {Soltan-Panahi},
  \citenamefont {Becker}, \citenamefont {Sengstock},\ and\ \citenamefont
  {Lewenstein}}]{Eckardt:2010}%
  \BibitemOpen
  \bibfield  {author} {\bibinfo {author} {\bibfnamefont {A.}~\bibnamefont
  {Eckardt}}, \bibinfo {author} {\bibfnamefont {P.}~\bibnamefont {Hauke}},
  \bibinfo {author} {\bibfnamefont {P.}~\bibnamefont {Soltan-Panahi}}, \bibinfo
  {author} {\bibfnamefont {C.}~\bibnamefont {Becker}}, \bibinfo {author}
  {\bibfnamefont {K.}~\bibnamefont {Sengstock}}, \ and\ \bibinfo {author}
  {\bibfnamefont {M.}~\bibnamefont {Lewenstein}},\ }\href {\doibase
  10.1209/0295-5075/89/10010} {\bibfield  {journal} {\bibinfo  {journal}
  {Europhysics Letters}\ }\textbf {\bibinfo {volume} {89}},\ \bibinfo {pages}
  {10010} (\bibinfo {year} {2010})}\BibitemShut {NoStop}%
\bibitem [{\citenamefont {Kosior}\ and\ \citenamefont
  {Sacha}(2014)}]{KosiorA2014}%
  \BibitemOpen
  \bibfield  {author} {\bibinfo {author} {\bibfnamefont {A.}~\bibnamefont
  {Kosior}}\ and\ \bibinfo {author} {\bibfnamefont {K.}~\bibnamefont {Sacha}},\
  }\href {\doibase 10.1103/PhysRevLett.112.045302} {\bibfield  {journal}
  {\bibinfo  {journal} {Physical Review Letters}\ }\textbf {\bibinfo {volume}
  {112}},\ \bibinfo {pages} {45302} (\bibinfo {year} {2014})}\BibitemShut
  {NoStop}%
\bibitem [{\citenamefont {Parker}\ \emph {et~al.}(2013)\citenamefont {Parker},
  \citenamefont {Ha},\ and\ \citenamefont {Chin}}]{ParkerC2013}%
  \BibitemOpen
  \bibfield  {author} {\bibinfo {author} {\bibfnamefont {C.~V.}\ \bibnamefont
  {Parker}}, \bibinfo {author} {\bibfnamefont {L.-C.}\ \bibnamefont {Ha}}, \
  and\ \bibinfo {author} {\bibfnamefont {C.}~\bibnamefont {Chin}},\ }\href
  {\doibase 10.1038/nphys2789} {\bibfield  {journal} {\bibinfo  {journal}
  {Nature Physics}\ }\textbf {\bibinfo {volume} {9}},\ \bibinfo {pages} {769}
  (\bibinfo {year} {2013})}\BibitemShut {NoStop}%
\bibitem [{\citenamefont {Kock}\ \emph {et~al.}(2015)\citenamefont {Kock},
  \citenamefont {{\"{O}}lschl{\"{a}}ger}, \citenamefont {Ewerbeck},
  \citenamefont {Huang}, \citenamefont {Mathey},\ and\ \citenamefont
  {Hemmerich}}]{KockT2015}%
  \BibitemOpen
  \bibfield  {author} {\bibinfo {author} {\bibfnamefont {T.}~\bibnamefont
  {Kock}}, \bibinfo {author} {\bibfnamefont {M.}~\bibnamefont
  {{\"{O}}lschl{\"{a}}ger}}, \bibinfo {author} {\bibfnamefont {A.}~\bibnamefont
  {Ewerbeck}}, \bibinfo {author} {\bibfnamefont {W.-M.}\ \bibnamefont {Huang}},
  \bibinfo {author} {\bibfnamefont {L.}~\bibnamefont {Mathey}}, \ and\ \bibinfo
  {author} {\bibfnamefont {A.}~\bibnamefont {Hemmerich}},\ }\href {\doibase
  10.1103/PhysRevLett.114.115301} {\bibfield  {journal} {\bibinfo  {journal}
  {Physical Review Letters}\ }\textbf {\bibinfo {volume} {114}},\ \bibinfo
  {pages} {115301} (\bibinfo {year} {2015})}\BibitemShut {NoStop}%
\bibitem [{\citenamefont {Wang}\ \emph {et~al.}(2022)\citenamefont {Wang},
  \citenamefont {Luo}, \citenamefont {Liu}, \citenamefont {Huang},
  \citenamefont {Li}, \citenamefont {Wu}, \citenamefont {Hemmerich},\ and\
  \citenamefont {Xu}}]{WangX2022}%
  \BibitemOpen
  \bibfield  {author} {\bibinfo {author} {\bibfnamefont {X.-Q.}\ \bibnamefont
  {Wang}}, \bibinfo {author} {\bibfnamefont {G.-Q.}\ \bibnamefont {Luo}},
  \bibinfo {author} {\bibfnamefont {J.-Y.}\ \bibnamefont {Liu}}, \bibinfo
  {author} {\bibfnamefont {G.-H.}\ \bibnamefont {Huang}}, \bibinfo {author}
  {\bibfnamefont {Z.-X.}\ \bibnamefont {Li}}, \bibinfo {author} {\bibfnamefont
  {C.}~\bibnamefont {Wu}}, \bibinfo {author} {\bibfnamefont {A.}~\bibnamefont
  {Hemmerich}}, \ and\ \bibinfo {author} {\bibfnamefont {Z.-F.}\ \bibnamefont
  {Xu}},\ }\href {http://arxiv.org/abs/2211.05578} {\bibfield  {journal}
  {\bibinfo  {journal} {arXiv:2211.05578}\ } (\bibinfo {year}
  {2022})}\BibitemShut {NoStop}%
\bibitem [{\citenamefont {Altman}\ and\ \citenamefont
  {Auerbach}(2002)}]{Altman:2002}%
  \BibitemOpen
  \bibfield  {author} {\bibinfo {author} {\bibfnamefont {E.}~\bibnamefont
  {Altman}}\ and\ \bibinfo {author} {\bibfnamefont {A.}~\bibnamefont
  {Auerbach}},\ }\href {\doibase 10.1103/PhysRevLett.89.250404} {\bibfield
  {journal} {\bibinfo  {journal} {Physical Review Letters}\ }\textbf {\bibinfo
  {volume} {89}},\ \bibinfo {pages} {250404} (\bibinfo {year}
  {2002})}\BibitemShut {NoStop}%
\bibitem [{\citenamefont {Yamamoto}\ \emph
  {et~al.}(2020{\natexlab{b}})\citenamefont {Yamamoto}, \citenamefont
  {Fukuhara},\ and\ \citenamefont {Danshita}}]{DYamamoto:2020}%
  \BibitemOpen
  \bibfield  {author} {\bibinfo {author} {\bibfnamefont {D.}~\bibnamefont
  {Yamamoto}}, \bibinfo {author} {\bibfnamefont {T.}~\bibnamefont {Fukuhara}},
  \ and\ \bibinfo {author} {\bibfnamefont {I.}~\bibnamefont {Danshita}},\
  }\href {\doibase 10.1038/s42005-020-0323-5} {\bibfield  {journal} {\bibinfo
  {journal} {Communications Physics}\ }\textbf {\bibinfo {volume} {3}},\
  \bibinfo {pages} {56} (\bibinfo {year} {2020}{\natexlab{b}})}\BibitemShut
  {NoStop}%
\bibitem [{\citenamefont {Keesling}\ \emph {et~al.}(2019)\citenamefont
  {Keesling}, \citenamefont {Omran}, \citenamefont {Levine}, \citenamefont
  {Bernien}, \citenamefont {Pichler}, \citenamefont {Choi}, \citenamefont
  {Samajdar}, \citenamefont {Schwartz}, \citenamefont {Silvi}, \citenamefont
  {Sachdev}, \citenamefont {Zoller}, \citenamefont {Endres}, \citenamefont
  {Greiner}, \citenamefont {Vuleti{\'{c}}},\ and\ \citenamefont
  {Lukin}}]{Keesling:2019}%
  \BibitemOpen
  \bibfield  {author} {\bibinfo {author} {\bibfnamefont {A.}~\bibnamefont
  {Keesling}}, \bibinfo {author} {\bibfnamefont {A.}~\bibnamefont {Omran}},
  \bibinfo {author} {\bibfnamefont {H.}~\bibnamefont {Levine}}, \bibinfo
  {author} {\bibfnamefont {H.}~\bibnamefont {Bernien}}, \bibinfo {author}
  {\bibfnamefont {H.}~\bibnamefont {Pichler}}, \bibinfo {author} {\bibfnamefont
  {S.}~\bibnamefont {Choi}}, \bibinfo {author} {\bibfnamefont {R.}~\bibnamefont
  {Samajdar}}, \bibinfo {author} {\bibfnamefont {S.}~\bibnamefont {Schwartz}},
  \bibinfo {author} {\bibfnamefont {P.}~\bibnamefont {Silvi}}, \bibinfo
  {author} {\bibfnamefont {S.}~\bibnamefont {Sachdev}}, \bibinfo {author}
  {\bibfnamefont {P.}~\bibnamefont {Zoller}}, \bibinfo {author} {\bibfnamefont
  {M.}~\bibnamefont {Endres}}, \bibinfo {author} {\bibfnamefont
  {M.}~\bibnamefont {Greiner}}, \bibinfo {author} {\bibfnamefont
  {V.}~\bibnamefont {Vuleti{\'{c}}}}, \ and\ \bibinfo {author} {\bibfnamefont
  {M.~D.}\ \bibnamefont {Lukin}},\ }\href {\doibase 10.1038/s41586-019-1070-1}
  {\bibfield  {journal} {\bibinfo  {journal} {Nature}\ }\textbf {\bibinfo
  {volume} {568}},\ \bibinfo {pages} {207} (\bibinfo {year}
  {2019})}\BibitemShut {NoStop}%
\bibitem [{\citenamefont {Goldman}\ and\ \citenamefont
  {Dalibard}(2014)}]{Goldman:2014}%
  \BibitemOpen
  \bibfield  {author} {\bibinfo {author} {\bibfnamefont {N.}~\bibnamefont
  {Goldman}}\ and\ \bibinfo {author} {\bibfnamefont {J.}~\bibnamefont
  {Dalibard}},\ }\href {\doibase 10.1103/PhysRevX.4.031027} {\bibfield
  {journal} {\bibinfo  {journal} {Physical Review X}\ }\textbf {\bibinfo
  {volume} {4}},\ \bibinfo {pages} {031027} (\bibinfo {year}
  {2014})}\BibitemShut {NoStop}%
\bibitem [{\citenamefont {Guo}\ \emph {et~al.}(2019)\citenamefont {Guo},
  \citenamefont {Zhang}, \citenamefont {Li}, \citenamefont {Shui},
  \citenamefont {Chen},\ and\ \citenamefont {Zhou}}]{Guo:2019}%
  \BibitemOpen
  \bibfield  {author} {\bibinfo {author} {\bibfnamefont {X.}~\bibnamefont
  {Guo}}, \bibinfo {author} {\bibfnamefont {W.}~\bibnamefont {Zhang}}, \bibinfo
  {author} {\bibfnamefont {Z.}~\bibnamefont {Li}}, \bibinfo {author}
  {\bibfnamefont {H.}~\bibnamefont {Shui}}, \bibinfo {author} {\bibfnamefont
  {X.}~\bibnamefont {Chen}}, \ and\ \bibinfo {author} {\bibfnamefont
  {X.}~\bibnamefont {Zhou}},\ }\href {\doibase 10.1364/oe.27.027786} {\bibfield
   {journal} {\bibinfo  {journal} {Optics Express}\ }\textbf {\bibinfo {volume}
  {27}},\ \bibinfo {pages} {27786} (\bibinfo {year} {2019})}\BibitemShut
  {NoStop}%
\bibitem [{\citenamefont {Arlinghaus}\ and\ \citenamefont
  {Holthaus}(2010)}]{Arlinghaus:2010}%
  \BibitemOpen
  \bibfield  {author} {\bibinfo {author} {\bibfnamefont {S.}~\bibnamefont
  {Arlinghaus}}\ and\ \bibinfo {author} {\bibfnamefont {M.}~\bibnamefont
  {Holthaus}},\ }\href {\doibase 10.1103/PhysRevA.81.063612} {\bibfield
  {journal} {\bibinfo  {journal} {Physical Review A}\ }\textbf {\bibinfo
  {volume} {81}},\ \bibinfo {pages} {063612} (\bibinfo {year}
  {2010})}\BibitemShut {NoStop}%
\bibitem [{\citenamefont {Martikainen}(2011)}]{Martikainen:2011}%
  \BibitemOpen
  \bibfield  {author} {\bibinfo {author} {\bibfnamefont {J.-P.}\ \bibnamefont
  {Martikainen}},\ }\href {\doibase 10.1103/PhysRevA.83.013610} {\bibfield
  {journal} {\bibinfo  {journal} {Physical Review A}\ }\textbf {\bibinfo
  {volume} {83}},\ \bibinfo {pages} {013610} (\bibinfo {year}
  {2011})}\BibitemShut {NoStop}%
\bibitem [{\citenamefont {Paul}\ and\ \citenamefont
  {Tiesinga}(2013)}]{Paul:2013}%
  \BibitemOpen
  \bibfield  {author} {\bibinfo {author} {\bibfnamefont {S.}~\bibnamefont
  {Paul}}\ and\ \bibinfo {author} {\bibfnamefont {E.}~\bibnamefont
  {Tiesinga}},\ }\href {\doibase 10.1103/PhysRevA.88.033615} {\bibfield
  {journal} {\bibinfo  {journal} {Physical Review A}\ }\textbf {\bibinfo
  {volume} {88}},\ \bibinfo {pages} {033615} (\bibinfo {year}
  {2013})}\BibitemShut {NoStop}%
\end{thebibliography}
%

\end{document}